\documentclass[english]{article}
\usepackage[T1]{fontenc}
\usepackage[latin9]{inputenc}
\usepackage{amsmath}
\usepackage{amssymb}
\usepackage{graphicx}
\usepackage{setspace}
\usepackage{esint}

\makeatletter
\newcommand{\lyxaddress}[1]{
\par {\raggedright #1
\vspace{1.4em}
\noindent\par}
}

\makeatother

\usepackage{babel}
\begin{document}

\title{\textbf{A formally exact master equation for open quantum systems
}\thanks{This is a first draft of the manuscript. More physical applications
of the master equation are being written up.}}

\author{Li Yu$^{1}$ and Eric J. Heller$^{1\text{,}2}$}
\maketitle
\begin{singlespace}

\lyxaddress{\begin{center}
\emph{$^{1}$Department of Physics, Harvard University, Cambridge,
MA 02138, USA}
\par\end{center}}

\lyxaddress{\begin{center}
\emph{$^{2}$Department of Chemistry and Chemical Biology, Harvard
University, Cambridge, MA 02138, USA}
\par\end{center}}
\end{singlespace}
\begin{abstract}
We present a succinct and intuitive derivation of a formally exact
master equation for general open quantum systems, without the use
of an ``inverse'' map which was invoked in previous works on formally
exact master equations. This formalism is applicable to non-Markovian
regimes. We derive a second-order equation of motion for the illustrative
spin-boson model at arbitrary temperatures, observing non-exponential
decoherence and relaxation. Limiting our generic derivation to zero
temperature, we also reproduce the result for the special case of
a vacuum bath in Phys. Rev. A 81, 042103 (2010).
\end{abstract}

\section{Introduction}

\begin{singlespace}
A closed quantum system does not interact with external quantum degrees
of freedom and its unitary dynamics is describable by the von Neumann
equation, \cite{breuer book} 
\begin{equation}
\frac{d}{dt}\rho_{total}(t)=-i\left[H_{total}(t),\rho_{total}(t)\right].
\end{equation}
An open quantum system interacts with external quantum degrees of
freedom (the ``environment''). \cite{breuer book} There have been
extensive studies on open quantum systems. \cite{breuer book,schlosshauer book}
It is well known that open quantum systems generally do not evolve
unitarily and the dynamics of their reduced density matrix $\rho_{S}(t)\equiv Tr_{E}\left(\rho_{total}(t)\right)$
cannot be adequately described by the von Neumann equation; the scope
of our work falls under the master equation approach to open system
dynamics, which goes beyond the von Neumann equation and aims to describe
non-unitary dynamics. \cite{breuer book}

Broadly speaking, quantum coherence plays an essential role in the
field of quantum information \cite{nielson chuang book} and quantum
control \cite{brumer book}. The loss of quantum coherence, or ``decoherence'',
generally arises in open systems, resulting from their interaction
with the environment. \cite{schlosshauer book} One focus of open
system study is thus on the decoherence aspect, besides other issues
such as dissipation. \cite{schlosshauer book} 

There have been much works on approximate approaches to open quantum
systems, \cite{breuer book} such as the widely used Born approximation
and Markovian approximation \cite{schlosshauer book,yamaguchi,Elenewski,jeske,kirton}.
However, from a theoretical point of view, these approximate approaches
do not adequately reveal the ``properties'' of open quantum system
dynamics. A formally exact master approach makes it possible to gain
insights into the ``properties'' of general open quantum system
dynamics, exact to every perturbative order. The closed form of the
equation of motion may already provide insights into the nature of
such dynamics, without it necessarily being solved. On the practical
side, these approximations may be unjustified in and inapplicable
to important situations. For example, the Markovian description does
not apply to various scenarios of physical, chemical, and/or biological
interest. \cite{aspuru,liang,fainberg} In principle, a formally exact
approach makes it possible to go beyond such restrictions and be more
widely applicable, including to non-Markovian regimes.

Formally exact approaches to general open quantum system dynamics
are studied in \cite{breuer,breuer book} with the time-convolutionless
projection operator technique. (See \cite{shibata 1977,shibata 1979,shibata 1980}
for the first proposal of this technique by Shibata et al.)  There
are also works outside of the field of open quantum systems, but on
formally exact approaches to average dynamics of closed quantum systems
\cite{james1,james2}. All the aforementioned formally exact approaches
\cite{breuer,breuer book,james1,james2} explicitly invoke some ``inverse''
in the derivations. Here we hope to dispense with the use of ``inverse''
in our derivation of the formally exact master equation. Our approach
will be direct and ``by construction'', rather than starting with
some ansatz.

Besides, exact master equations are constructed in \cite{vacchini}
for a two-level system decaying to a bath initially in vacuum state,
wherein various techniques including the time-convolutionless method
are discussed. The works \cite{gaussian 1,gaussian 2} present exact
master equations for the case of Gaussian open quantum system dynamics.
An exact master equation for quantum Brownian motion is presented
in \cite{hu} with the influence functional method. The work in \cite{tu}
shows an exact master equation for electrons in double dot by extending
the influence functional method to fermionic environments. There is
also a work on post-Markovian master equation through a measurement
approach \cite{shabani}. 
\end{singlespace}

It is our goal to provide a succinct and intuitive, and yet sound,
approach to deriving a formally exact master equation for general
open quantum systems, that is, without restrictions on the type of
system, environment, or system-environment interaction. This is the
subject of Section 2 of this paper. The formalism developed in Section
2 is then applied to study the spin-boson model as in Section 3 to
illustrate the use of the master equation.
\begin{singlespace}

\section{Theory}
\end{singlespace}
\begin{singlespace}

\subsection{Derivations}
\end{singlespace}
\begin{singlespace}

\subsubsection*{Series expansion of full dynamics}
\end{singlespace}

\begin{singlespace}
We start with the equation of motion for the full system-environment
dynamics, \footnote{Throughout the paper we formally set $\hbar=1$ for notational convenience
unless otherwise noted.} 
\begin{equation}
i\frac{d}{dt}\rho_{SE}(t)=\left[H_{SE}(t),\rho_{SE}(t)\right],
\end{equation}
where $H_{SE}(t)$ and $\rho_{SE}(t)$ are the interaction Hamiltonian
and the full system-environment density matrix in the interaction
picture respectively. \cite{shankar book,breuer book} Following a
standard approach to parametrize the Hamiltonian $H_{SE}(t)$ by $\lambda$,
\cite{james1,james2} we have 
\begin{equation}
i\frac{d}{dt}\rho_{SE}(t)=\lambda\left[H_{SE}(t),\rho_{SE}(t)\right].
\end{equation}
The full density matrix $\rho_{SE}(t)$ evolves unitarily, 
\begin{equation}
\rho_{SE}(t)=U(t,0)\rho_{SE}(0)U^{\dagger}(t,0),
\end{equation}
where the uniraty operator $U(t,0)$ obeys the equation of motion
\begin{equation}
i\frac{d}{dt}U(t,0)=\lambda H_{SE}(t)U(t,0).\label{eq:1. U eom}
\end{equation}

We suppose the unitary operator can be expanded in a power series
of $\lambda$: \cite{shankar book,james1,james2}
\begin{equation}
U(t,0)=\sum_{n=0}^{\infty}\lambda^{n}U_{n}(t,0).\label{eq:1. U expansion}
\end{equation}
Plugging Eq.(\ref{eq:1. U expansion}) into Eq.(\ref{eq:1. U eom}),
we have 
\begin{eqnarray}
i\frac{d}{dt}U_{0}(t,0) & = & 0,\\
i\frac{d}{dt}U_{n}(t,0) & = & H_{SE}(t)U_{n-1}(t,0)\;\;\left(n=1,2,....\right).
\end{eqnarray}
Solving the above equations, we have
\begin{eqnarray}
U_{0}(t,0) & = & \mathbb{I},\label{eq:1. U0}\\
U_{n}(t,0) & = & -i\,\intop_{0}^{t}dt'H_{SE}(t')U_{n-1}(t',0)\;\;\left(n=1,2,....\right).\label{eq:1. Un}
\end{eqnarray}

The full system-environment dynamics can thus be expressed as 
\begin{equation}
\rho_{SE}(t)=\sum_{m=0}^{\infty}\sum_{n=0}^{\infty}\lambda^{m+n}U_{m}(t,0)\rho_{S}(0)\otimes\rho_{E}(0)U_{n}^{\dagger}(t,0).
\end{equation}

\end{singlespace}
\begin{singlespace}

\subsubsection*{Reduced dynamics}
\end{singlespace}

\begin{singlespace}
The reduced density matrix of the system is the partial trace of the
full density matrix over environmental degrees of freedom \cite{breuer book,schlosshauer book}
\begin{equation}
\rho_{S}(t)=Tr_{E}\left(\rho_{SE}(t)\right)=\sum_{m=0}^{\infty}\sum_{n=0}^{\infty}\lambda^{m+n}Tr_{E}\left(U_{m}(t,0)\rho_{S}(0)\otimes\rho_{E}(0)U_{n}^{\dagger}(t,0)\right).
\end{equation}
For convenience in subsequent derivation, let's re-write the mapping
as 
\begin{eqnarray}
\rho_{S}(t) & = & \sum_{k=0}^{\infty}\lambda^{k}\mathfrak{\mathcal{E}}_{k,t}\left(\rho_{S}(0)\right)\nonumber \\
 & = & \rho_{S}(0)+\sum_{k=1}^{\infty}\lambda^{k}\mathfrak{\mathcal{E}}_{k,t}\left(\rho_{S}(0)\right)\nonumber \\
 & \equiv & \left(\mathbb{I}+\mathfrak{\mathcal{E}}_{t}\right)\left(\rho_{S}(0)\right),
\end{eqnarray}
where
\begin{eqnarray}
\mathfrak{\mathcal{E}}_{t}\left(\rho\right) & \equiv & \sum_{k=1}^{\infty}\lambda^{k}\mathfrak{\mathcal{E}}_{k,t}\left(\rho\right),\label{eq:1. eps total}\\
\mathfrak{\mathcal{E}}_{k,t}\left(\rho\right) & \equiv & \sum_{j=0}^{k}Tr_{E}\left(U_{k-j}(t,0)\rho\otimes\rho_{E}(0)U_{j}^{\dagger}(t,0)\right)\;\;\left(k=1,2,....\right).\label{eq:1. eps k}
\end{eqnarray}

Note that because $\mathfrak{\mathcal{E}}_{t}\left(\rho\right)=\sum_{k=1}^{\infty}\lambda^{k}\mathfrak{\mathcal{E}}_{k,t}\left(\rho\right)\sim\mathcal{O}\left(\lambda\right)$,
we know $\mathfrak{\mathcal{E}}_{t}\left(\rho\right)$ approaches
zero as $\lambda\rightarrow0$. Also, by definition, $\mathfrak{\mathcal{E}}_{t}\left(\rho\right)$
approaches zero as $t\rightarrow0$.
\end{singlespace}
\begin{singlespace}

\subsubsection*{The $Y_{N,t}$ map}
\end{singlespace}

\begin{singlespace}
The key to obtaining a formally exact, time-local equation of motion
in closed form is the following step. Let's define a linear map central
to our construction: 
\begin{equation}
Y_{N,t}\left(\rho\right)\equiv\sum_{n=0}^{N}(-1)^{n}\mathfrak{\mathcal{E}}_{t}^{(n)}\left(\rho\right),\label{eq:1. Y map}
\end{equation}
where $\mathfrak{\mathcal{E}}_{t}^{(n)}\left(\rho\right)\equiv\mathfrak{\mathcal{E}}_{t}\left(\mathfrak{\mathcal{E}}_{t}\left(...\mathfrak{\mathcal{E}}_{t}\left(\rho\right)\right)\right)$
is a composition of $n$ $\mathfrak{\mathcal{E}}_{t}$ maps. Then,
applying this linear map to the system's density matrix at time t
yields
\begin{eqnarray}
Y_{N,t}\left(\rho_{S}(t)\right) & = & \sum_{n=0}^{N}(-1)^{n}\mathfrak{\mathcal{E}}_{t}^{(n)}\left(\left(\mathbb{I}+\mathfrak{\mathcal{E}}_{t}\right)\left(\rho_{S}(0)\right)\right)\nonumber \\
 & = & \mathbb{I}\left(\left(\mathbb{I}+\mathfrak{\mathcal{E}}_{t}\right)\left(\rho_{S}(0)\right)\right)-\mathfrak{\mathcal{E}}_{t}\left(\left(\mathbb{I}+\mathfrak{\mathcal{E}}_{t}\right)\left(\rho_{S}(0)\right)\right)\nonumber \\
 &  & +\mathfrak{\mathcal{E}}_{t}\left(\mathfrak{\mathcal{E}}_{t}\left(\left(\mathbb{I}+\mathfrak{\mathcal{E}}_{t}\right)\left(\rho_{S}(0)\right)\right)\right)-....\nonumber \\
 & = & \mathbb{I}\left(\rho_{S}(0)\right)+\mathfrak{\mathcal{E}}_{t}\left(\rho_{S}(0)\right)-\mathfrak{\mathcal{E}}_{t}\left(\rho_{S}(0)\right)-\mathfrak{\mathcal{E}}_{t}\left(\mathfrak{\mathcal{E}}_{t}\left(\rho_{S}(0)\right)\right)\nonumber \\
 &  & +\mathfrak{\mathcal{E}}_{t}\left(\mathfrak{\mathcal{E}}_{t}\left(\rho_{S}(0)\right)\right)+\mathfrak{\mathcal{E}}_{t}\left(\mathfrak{\mathcal{E}}_{t}\left(\mathfrak{\mathcal{E}}_{t}\left(\rho_{S}(0)\right)\right)\right)-....\nonumber \\
 & = & \left(\mathbb{I}+(-1)^{N}\mathfrak{\mathcal{E}}_{t}^{(N+1)}\right)\left(\rho_{S}(0)\right).\label{eq:1. Yn}
\end{eqnarray}
Denoting $\rho_{S}(t)\equiv\rho_{t}$ and $\rho_{S}(0)\equiv\rho_{0}$
for notational convenience, we now have the key equality in our work:
\begin{eqnarray}
\rho_{0} & = & Y_{N,t}\left(\rho_{t}\right)+(-1)^{N+1}\mathfrak{\mathcal{E}}_{t}^{(N+1)}\left(\rho_{0}\right)\nonumber \\
 & = & \sum_{n=0}^{N}(-1)^{n}\mathfrak{\mathcal{E}}_{t}^{(n)}\left(\rho_{t}\right)+(-1)^{N+1}\mathfrak{\mathcal{E}}_{t}^{(N+1)}\left(\rho_{0}\right).\label{eq:1. Y invert}
\end{eqnarray}
What it does is to express the initial system's state $\rho_{0}$
in terms of the system's state at time t $\rho_{t}$ (with a residual
term $(-1)^{N+1}\mathfrak{\mathcal{E}}_{t}^{(N+1)}\left(\rho_{0}\right)$
that can be neglected to certain perturbative orders).

Note that the form of the $Y_{N,t}\left(\rho\right)$ map might bear
some resemblence to the $\left[1-\varSigma(t)\right]^{-1}$ super-operator
in \cite{breuer,breuer book}, but there is at least one important
difference besides others: here we make no use of an inverse map,
whereas \cite{breuer,breuer book} assumes an inverse.

Our work might be mathematically equivalent to the apparently different
work in \cite{breuer,breuer book}, wherein more complicated theoretical
constructs are used, such as the projection operator technique and
antichronological time-ordering. In fact, any formulation of a general
exact master equation should be mathematically equivalent to any other
formulation in every order of the perturbative parameter $\lambda$.
In any case, our work is independently constructed, with all the derivation
steps naturally motivated and intermediate terms intuitively defined.
It is our goal to formulate a succint and intuitive, and yet sound,
approach to deriving a formally exact master equation for general
open quantum systems, and we believe that we have achieved this goal
with our work.
\end{singlespace}
\begin{singlespace}

\subsubsection*{Equation of motion}
\end{singlespace}

\begin{singlespace}
Taking the time derivative of the system's reduced density matrix
and making use of Eq.(\ref{eq:1. Y invert}), we have
\begin{eqnarray}
\frac{d}{dt}\rho_{t} & = & \frac{d}{dt}\left(\mathbb{I}+\mathfrak{\mathcal{E}}_{t}\right)\left(\rho_{0}\right)\nonumber \\
 & = & \dot{\mathfrak{\mathcal{E}_{t}}}\left(\rho_{0}\right)\nonumber \\
 & = & \dot{\mathfrak{\mathcal{E}_{t}}}\left(\sum_{n=0}^{N}(-1)^{n}\mathfrak{\mathcal{E}}_{t}^{(n)}\left(\rho_{t}\right)+(-1)^{N+1}\mathfrak{\mathcal{E}}_{t}^{(N+1)}\left(\rho_{0}\right)\right).
\end{eqnarray}
Note that in the third equality the $Y_{N,t}\left(\rho\right)$ map
as in Eqs.(\ref{eq:1. Y map}, \ref{eq:1. Y invert}) has done the
crucial job of re-expressing the right-hand side of the equation in
terms of the quantity of interest, namely the system's state at time
t $\rho_{t}$. Therefore, we have
\begin{equation}
\frac{d}{dt}\rho_{t}=\sum_{n=0}^{N}(-1)^{n}\dot{\mathfrak{\mathcal{E}_{t}}}\left(\mathfrak{\mathcal{E}}_{t}^{(n)}\left(\rho_{t}\right)\right)+(-1)^{N+1}\dot{\mathfrak{\mathcal{E}_{t}}}\left(\mathfrak{\mathcal{E}}_{t}^{(N+1)}\left(\rho_{0}\right)\right).\label{eq:1. eom inhomo}
\end{equation}
Note that, up to this point, no approximation has been made and Eq.(\ref{eq:1. eom inhomo})
is formally exact. 

With Eq.(\ref{eq:1. eom inhomo}), we can systematically make approximations,
that is, collecting like-order terms in $\lambda$ and truncating
the series as needed. Since $\mathfrak{\mathcal{E}}_{t}\left(\rho\right)\sim\mathcal{O}\left(\lambda\right)\Rightarrow\dot{\mathfrak{\mathcal{E}}_{t}}\left(\mathfrak{\mathcal{E}}_{t}^{(N+1)}\left(\rho_{0}\right)\right)\sim\mathcal{O}\left(\lambda^{N+2}\right)$,
if we want to consider $M$th-order approximation, we can always choose
$N\geqslant M-1$, so that the residual term $(-1)^{N+1}\dot{\mathfrak{\mathcal{E}_{t}}}\left(\mathfrak{\mathcal{E}}_{t}^{(N+1)}\left(\rho_{0}\right)\right)$
may be neglected in our intended approximation and thus its presence
in Eq.(\ref{eq:1. eom inhomo}) does not matter. \footnote{Loosely speaking, in order for the residual term $(-1)^{N+1}\dot{\mathfrak{\mathcal{E}_{t}}}\left(\mathfrak{\mathcal{E}}_{t}^{(N+1)}\left(\rho_{0}\right)\right)$
to be negligible compared to lower order terms like $(-1)^{N}\dot{\mathfrak{\mathcal{E}_{t}}}\left(\mathfrak{\mathcal{E}}_{t}^{(N)}\left(\rho_{t}\right)\right)$
in Eq.(\ref{eq:1. eom inhomo}), it apparently requires the map $\mathfrak{\mathcal{E}}_{t}\left(\ldots\right)$
be reasonably small. As we discuss earlier, the time-dependent map
$\mathfrak{\mathcal{E}}_{t}\left(\ldots\right)\rightarrow0$ as $t\rightarrow0$
by definition; also, as the coupling strength approaches zero, the
interaction Hamiltonian tends to vanish, thus $\mathfrak{\mathcal{E}}_{t}\left(\ldots\right)\rightarrow0$
as well. Therefore, our approximation should work in the short time
and/or weak coupling regimes. We do not extrapolate this approximation
to the long time or strong coupling regimes.}

Idealistically, we may hope to obtain a formally exact, time-local,
linear homogeneous differential equation as the equation of motion.
This can be formally achieved by taking the $N\rightarrow\infty$
limit on the right-hand side of Eq.(\ref{eq:1. eom inhomo}). Loosely
speaking, as $\lim_{N\rightarrow\infty}(-1)^{N+1}\dot{\mathfrak{\mathcal{E}}_{t}}\left(\mathfrak{\mathcal{E}}_{t}^{(N+1)}\left(\rho_{0}\right)\right)\sim\lim_{N\rightarrow\infty}\mathcal{O}\left(\lambda^{N+2}\right)\rightarrow0$,
the residual term may be dropped, and we have
\begin{equation}
\frac{d}{dt}\rho_{t}=\sum_{n=0}^{\infty}(-1)^{n}\dot{\mathfrak{\mathcal{E}_{t}}}\left(\mathfrak{\mathcal{E}}_{t}^{(n)}\left(\rho_{t}\right)\right),\label{eq:1. eom homo}
\end{equation}
which is formally a linear homogeneous differential equation, albeit
with infinitely many terms. \footnote{Implicit in this discussion is the convergence of the infinite series
in Eq.(\ref{eq:1. eom homo}). Loosely speaking, in order for the
infinite series to converge, it apparently requires the higher order
terms (i.e. $\dot{\mathfrak{\mathcal{E}_{t}}}\left(\mathfrak{\mathcal{E}}_{t}^{(n)}\left(\rho_{t}\right)\right)$
with larger $n$) be progressively smaller. As we discuss earlier,
$\mathfrak{\mathcal{E}}_{t}\left(\ldots\right)\rightarrow0$ as $t\rightarrow0$;
also, $\mathfrak{\mathcal{E}}_{t}\left(\ldots\right)\rightarrow0$
as coupling approaches zero. As $\mathfrak{\mathcal{E}}_{t}\left(\ldots\right)\rightarrow0$,
$\dot{\mathfrak{\mathcal{E}_{t}}}\left(\mathfrak{\mathcal{E}}_{t}^{(n)}\left(\rho_{t}\right)\right)$
should be progressively smaller for larger $n$, thus our discussion
should be valid in the short time and/or weak coupling regimes. We
do not extrapolate this discussion to the long time or strong coupling
regimes.} However, note that this $N\rightarrow\infty$ formal treatment and
the resulting linear homogeneous differential equation are not necessary
for obtaining an $M$th-order approximate equation of motion for the
system's reduced dynamics, the latter of which is all that matters
in applications. In other words, this $N\rightarrow\infty$ formal
treatment can be dispensed with no practical implications.
\end{singlespace}
\begin{singlespace}

\subsection{Second-order equation of motion}
\end{singlespace}

\begin{singlespace}
In many cases, one is interested in the second-order approximate equation
of motion, as it is usually the leading order term that exhibits interesting
effects such as decoherence. For second-order approximation, let $N=2-1=1$
in Eq.(\ref{eq:1. eom inhomo}):
\begin{eqnarray}
\frac{d}{dt}\rho_{t} & = & \dot{\mathfrak{\mathcal{E}_{t}}}\left(\rho_{t}\right)-\dot{\mathfrak{\mathcal{E}_{t}}}\left(\mathfrak{\mathcal{E}}_{t}\left(\rho_{t}\right)\right)+\mathcal{O}\left(\lambda^{3}\right)\nonumber \\
 & = & \left[\lambda\dot{\mathfrak{\mathcal{E}}_{1,t}}\left(\rho_{t}\right)+\lambda^{2}\dot{\mathfrak{\mathcal{E}}_{2,t}}\left(\rho_{t}\right)+....\right]\nonumber \\
 &  & -\left[\left(\lambda\dot{\mathfrak{\mathcal{E}}_{1,t}}+\lambda^{2}\dot{\mathfrak{\mathcal{E}}_{2,t}}+....\right)\left(\lambda\mathfrak{\mathcal{E}}_{1,t}\left(\rho_{t}\right)+\lambda^{2}\dot{\mathfrak{\mathcal{E}_{2,t}}}\left(\rho_{t}\right)+....\right)\right]\nonumber \\
 &  & +\mathcal{O}\left(\lambda^{3}\right)\nonumber \\
 & = & \lambda\dot{\mathfrak{\mathcal{E}}_{1,t}}\left(\rho_{t}\right)+\lambda^{2}\left[\dot{\mathfrak{\mathcal{E}}_{2,t}}\left(\rho_{t}\right)-\dot{\mathfrak{\mathcal{E}}_{1,t}}\left(\mathfrak{\mathcal{E}}_{1,t}\left(\rho_{t}\right)\right)\right]+\mathcal{O}\left(\lambda^{3}\right).
\end{eqnarray}
Therefore, the second-order equation of motion is 
\begin{equation}
\frac{d}{dt}\rho_{t}=\mathcal{L}_{1,t}\left(\rho_{t}\right)+\mathcal{L}_{2,t}\left(\rho_{t}\right),\label{eq:1. 2nd-order eom}
\end{equation}
where $\mathcal{L}_{1,t}\left(\rho\right)$ and $\mathcal{L}_{2,t}\left(\rho\right)$
are defined for an arbitrary $\rho$ as
\begin{eqnarray}
\mathcal{L}_{1,t}\left(\rho\right) & = & \dot{\mathfrak{\mathcal{E}}_{1,t}}\left(\rho\right),\\
\mathcal{L}_{2,t}\left(\rho\right) & = & \dot{\mathfrak{\mathcal{E}}_{2,t}}\left(\rho\right)-\dot{\mathfrak{\mathcal{E}}_{1,t}}\left(\mathfrak{\mathcal{E}}_{1,t}\left(\rho\right)\right).
\end{eqnarray}

More specifically, we can work out the formal expressions of $\mathcal{L}_{1,t}\left(\rho\right)$
and $\mathcal{L}_{2,t}\left(\rho\right)$ in terms of $H_{SE}(t)$
and $\rho_{E0}$:
\begin{eqnarray}
\mathcal{L}_{1,t}\left(\rho\right) & = & \dot{\mathfrak{\mathcal{E}}_{1,t}}\left(\rho\right)\nonumber \\
 & = & Tr_{E}\left[\dot{U}_{1}(t,0)\rho\otimes\rho_{E0}\right]+Tr_{E}\left[\rho\otimes\rho_{E0}\dot{U}_{1}^{\dagger}(t,0)\right]\nonumber \\
 & = & -i\left(Tr_{E}\left[H_{SE}(t)\rho\otimes\rho_{E0}\right]-Tr_{E}\left[\rho\otimes\rho_{E0}H_{SE}(t)\right]\right);
\end{eqnarray}
\begin{eqnarray}
\mathcal{L}_{2,t}\left(\rho\right) & = & \dot{\mathfrak{\mathcal{E}}_{2,t}}\left(\rho\right)-\dot{\mathfrak{\mathcal{E}}_{1,t}}\left(\mathfrak{\mathcal{E}}_{1,t}\left(\rho\right)\right)\nonumber \\
 & = & Tr_{E}\{\:\dot{U}_{2}(t,0)\rho\otimes\rho_{E0}+\dot{U}_{1}(t,0)\rho\otimes\rho_{E0}U_{1}^{\dagger}(t,0)\nonumber \\
 &  & +U_{1}(t,0)\rho\otimes\rho_{E0}\dot{U}_{1}^{\dagger}(t,0)+\rho\otimes\rho_{E0}\dot{U}_{2}^{\dagger}(t,0)\:\}\nonumber \\
 &  & +i\,Tr_{E}\{\:H_{SE}(t)\mathfrak{\mathcal{E}}_{1,t}\left(\rho\right)\otimes\rho_{E0}-\mathfrak{\mathcal{E}}_{1,t}\left(\rho\right)\otimes\rho_{E0}H_{SE}(t)\:\}\nonumber \\
 & = & -\int_{0}^{t}dt'Tr_{E}\{\:H_{SE}(t)H_{SE}(t')\rho\otimes\rho_{E0}-H_{SE}(t)\rho\otimes\rho_{E0}H_{SE}(t')\nonumber \\
 &  & -H_{SE}(t')\rho\otimes\rho_{E0}H_{SE}(t)+\rho\otimes\rho_{E0}H_{SE}(t')H_{SE}(t)\:\}\nonumber \\
 &  & +\int_{0}^{t}dt'Tr_{E}\{\:H_{SE}(t)\left(Tr_{E}\left[H_{SE}(t')\rho\otimes\rho_{E0}-\rho\otimes\rho_{E0}H_{SE}(t')\right]\right)\otimes\rho_{E0}\nonumber \\
 &  & -\left(Tr_{E}\left[H_{SE}(t')\rho\otimes\rho_{E0}-\rho\otimes\rho_{E0}H_{SE}(t')\right]\right)\otimes\rho_{E0}H_{SE}(t)\:\}.
\end{eqnarray}

In general, the interaction Hamiltonian $H_{SE}(t)$ can be expressed
in terms of operators on the system Hilbert space $\left\{ S_{n}(t)\right\} $
and those on the bath Hilbert space $\left\{ E_{n}(t)\right\} $ as
\cite{breuer book}
\begin{equation}
H_{SE}(t)=\sum_{n}S_{n}(t)\otimes E_{n}(t).\label{eq:1. product form}
\end{equation}
With this, $\mathcal{L}_{1,t}\left(\rho\right)$ and $\mathcal{L}_{2,t}\left(\rho\right)$
can be re-expressed as:
\begin{eqnarray}
\mathcal{L}_{1,t}\left(\rho\right) & = & -i\left(\sum_{n}Tr_{E}\left(S_{n}(t)\rho\otimes E_{n}(t)\rho_{E0}\right)-\sum_{n}Tr_{E}\left(\rho S_{n}(t)\otimes\rho_{E0}E_{n}(t)\right)\right)\nonumber \\
 & = & -i\sum_{n}Tr_{E}\left(\rho_{E0}E_{n}(t)\right)\left[S_{n}(t),\:\rho\right],\label{eq:1. L1}
\end{eqnarray}
\begin{eqnarray}
\mathcal{L}_{2,t}\left(\rho\right) & = & -\int_{0}^{t}dt'\sum_{m}\sum_{n}\left(Tr_{E}\left(\rho_{E0}E_{m}(t)E_{n}(t')\right)-Tr_{E}\left(\rho_{E0}E_{m}(t)\right)Tr_{E}\left(\rho_{E0}E_{n}(t')\right)\right)\nonumber \\
 &  & \left[S_{m}(t),\:S_{n}(t')\rho\right]\nonumber \\
 &  & +\int_{0}^{t}dt'\sum_{m}\sum_{n}\left(Tr_{E}\left(\rho_{E0}E_{n}(t')E_{m}(t)\right)-Tr_{E}\left(\rho_{E0}E_{n}(t')\right)Tr_{E}\left(\rho_{E0}E_{m}(t)\right)\right)\nonumber \\
 &  & \left[S_{m}(t),\:\rho S_{n}(t')\right].\label{eq:1. L2}
\end{eqnarray}

\end{singlespace}
\begin{singlespace}

\subsubsection*{Main result}
\end{singlespace}

\begin{singlespace}
In summary, for an open quantum system interacting with a bath via
the Hamiltonian $H_{SE}(t)=\sum_{n}S_{n}(t)\otimes E_{n}(t)$, the
initial state of the bath being $\rho_{E0}$, the equation of motion
for the system's reduced density matrix $\rho_{t}$ is (up to second
order) 
\begin{equation}
\frac{d}{dt}\rho_{t}=-i\,\left[H_{eff}(t),\:\rho_{t}\right]+\mathcal{L}_{2,t}\left(\rho_{t}\right),\label{eq:1. 2nd-order eom re-written}
\end{equation}
where the first-order effective Hamiltonian is
\begin{equation}
H_{eff}(t)\equiv\sum_{n}Tr_{E}\left(\rho_{E0}E_{n}(t)\right)S_{n}(t),\label{eq:1. L1 eom}
\end{equation}
and the second-order term is
\begin{equation}
\mathcal{L}_{2,t}\left(\rho\right)=-\sum_{m}\sum_{n}\int_{0}^{t}dt'\left(\mathcal{C}_{mn}(t,t')\left[S_{m}(t),\:S_{n}(t')\rho\right]-\mathcal{C}_{nm}(t',t)\left[S_{m}(t),\:\rho S_{n}(t')\right]\right),\label{eq:1. L2 eom}
\end{equation}
with the coefficients being 
\begin{equation}
\mathcal{C}_{jk}(t,t')\equiv Tr_{E}\left(\rho_{E0}E_{j}(t)E_{k}(t')\right)-Tr_{E}\left(\rho_{E0}E_{j}(t)\right)Tr_{E}\left(\rho_{E0}E_{k}(t')\right).\label{eq:1. L2 coeff}
\end{equation}

\end{singlespace}

Second-order non-Markovian master equations like this are previously
studied in the literature. For example, \cite{breuer,breuer book}
shows a time-convolutionless projection operator approach, wherein
Eqs.(9.52, 9.61) of Ref.\cite{breuer book} is a second-order non-Markovian
master equation, though with the first-order effective Hamiltonian
vanishing due to the vanishing odd moments of the interaction Hamiltonian
with respect to the environmental state. 
\begin{singlespace}

\subsection{Higher-order equations of motion}
\end{singlespace}

\begin{singlespace}
With the master equation formalism developed herein, one can systematically
investigate an open quantum system's dynamics to higher orders. For
example, if one is interested in the reduced dynamics up to $M$-th
order, one can first set $N=M-1$ in Eq.(\ref{eq:1. eom inhomo})
to obtain 
\begin{eqnarray}
\frac{d}{dt}\rho_{t} & = & \sum_{n=0}^{M-1}(-1)^{n}\dot{\mathfrak{\mathcal{E}_{t}}}\left(\mathfrak{\mathcal{E}}_{t}^{(n)}\left(\rho_{t}\right)\right)+(-1)^{M}\dot{\mathfrak{\mathcal{E}_{t}}}\left(\mathfrak{\mathcal{E}}_{t}^{(M)}\left(\rho_{0}\right)\right)\nonumber \\
 & = & \sum_{n=0}^{M-1}(-1)^{n}\dot{\mathfrak{\mathcal{E}_{t}}}\left(\mathfrak{\mathcal{E}}_{t}^{(n)}\left(\rho_{t}\right)\right)+\mathcal{O}\left(\lambda^{M+1}\right),\label{eq:1. eom higher order big O}
\end{eqnarray}
then work out the terms $\dot{\mathfrak{\mathcal{E}_{t}}}\left(\mathfrak{\mathcal{E}}_{t}^{(n)}\left(\rho_{t}\right)\right)$
according to Eqs.(\ref{eq:1. eps total}, \ref{eq:1. eps k}),
\begin{eqnarray}
\mathfrak{\mathcal{E}}_{t}\left(\rho\right) & = & \sum_{k=1}^{\infty}\lambda^{k}\mathfrak{\mathcal{E}}_{k,t}\left(\rho\right),\\
\mathfrak{\mathcal{E}}_{k,t}\left(\rho\right) & \equiv & \sum_{j=0}^{k}Tr_{E}\left(U_{k-j}(t,0)\rho\otimes\rho_{E}(0)U_{j}^{\dagger}(t,0)\right),
\end{eqnarray}
with $U_{n}(t,0)$ defined as in Eqs.(\ref{eq:1. U0},\ref{eq:1. Un}),
and then collect like order terms up to $M$-th order (dropping higher-order
contributions) to obtain an equation of the form 
\begin{equation}
\frac{d}{dt}\rho_{t}=-i\,\left[H_{eff}(t),\:\rho_{t}\right]+\mathcal{L}_{2,t}\left(\rho_{t}\right)+\sum_{n=3}^{M}\mathcal{L}_{n,t}\left(\rho_{t}\right),\label{eq:1. eom higher order}
\end{equation}
with every term $\mathcal{L}_{n,t}\left(\rho\right)$ in Eq.(\ref{eq:1. eom higher order})
well defined. All these steps can be carried out mechanically.
\end{singlespace}

Non-Markovian master equations of higher orders are also known in
the literature. See \cite{breuer,breuer book} again, for example,
wherein Eqs.(9.41, 9.42, 9.47, 9.51) of Ref.\cite{breuer book} show
some higher-order terms of the non-Markovian master equation. 

\section{Example: Spin-boson model}

\begin{singlespace}
A two-level system (TLS) interacting with bosonic field modes is extensively
studied and widely used in the open quantum systems literature. \cite{breuer book,schlosshauer book,yelin,breuer,liang,vacchini}
Here we will use the spin-boson model as an illustrative example for
the master equation formalism developed above.
\end{singlespace}
\begin{singlespace}

\subsection{Problem description}
\end{singlespace}

\begin{singlespace}
For a two-level system (TLS) interacting with a bosonic field, the
total Hamiltonian is (in Schrodinger picture) \cite{breuer book}
\begin{equation}
H_{total}=\frac{\omega_{0}}{2}\sigma_{z}+\sum_{k}\omega_{k}b_{k}^{\dagger}b_{k}+\sum_{k}g_{k}\left(\sigma_{+}b_{k}+\sigma_{-}b_{k}^{\dagger}\right),
\end{equation}
where the first term is the self-Hamiltonian of the TLS ($\omega_{0}$
being the energy spacing), the second term is the self-Hamiltonian
of a collection of independent bosonic modes ($b_{k}$ and $b_{k}^{\dagger}$
being the annihilation and creation operators of $k-th$ mode, $\omega_{k}$
being its frequency) \cite{breuer book}, and the third term is the
system-bath interaction ($g_{k}$ being the coupling strength between
TLS and $k-th$ field mode, and $\sigma_{+}$ leading to transition
from TLS's ground state to its excited state while $\sigma_{-}$ doing
the opposite) \cite{breuer book}.

Treating $H_{SE}=\sum_{k}g_{k}\left(\sigma_{+}b_{k}+\sigma_{-}b_{k}^{\dagger}\right)$
as a perturbation to the unperturbed Hamiltonian $H_{0}=\frac{\omega_{0}}{2}\sigma_{z}+\sum_{k}\omega_{k}b_{k}^{\dagger}b_{k}$
and switching to the interaction picture \cite{shankar book} (i.e.
the ``rotating frame'' generated by $H_{0}$), we have
\begin{equation}
H_{SE}^{(int-pic)}(t)=\sum_{k}g_{k}\left(\sigma_{+}b_{k}e^{-i(\omega_{k}-\omega_{0})t}+\sigma_{-}b_{k}^{\dagger}e^{i(\omega_{k}-\omega_{0})t}\right).\label{eq:int-H full}
\end{equation}
\footnote{Hereafter we drop the superscript ``interaction picture'' for notational
convenience and have in mind all operators are in the interaction
picture unless otherwise noted.}

Suppose the bosonic field is initially in the thermal state, that
is, 
\begin{equation}
\rho_{E0}=\frac{1}{Z}\exp\left(-\beta H_{field}\right),
\end{equation}
where $Z=Tr_{E}\left(\exp\left(-\beta H_{field}\right)\right)$ is
the partition function and $\beta=1/k_{B}T$ is the inverse temperature.
\cite{schlosshauer book,huang book,mandel book} In this example,
we have 
\begin{eqnarray}
\rho_{E0} & = & \prod_{k}\otimes\left(\frac{1}{Z_{k}}\sum_{m_{k}=0}^{\infty}e^{-m_{k}\beta\omega_{k}}|m_{k}\rangle\langle m_{k}|\right)\nonumber \\
 & = & \frac{1}{Z}\prod_{k}\otimes\left(\sum_{m_{k}=0}^{\infty}e^{-m_{k}\beta\omega_{k}}|m_{k}\rangle\langle m_{k}|\right),\label{eq:thermal state - product}
\end{eqnarray}
where $Z_{k}=\sum_{m_{k}=0}^{\infty}e^{-m_{k}\beta\omega_{k}}$ and
$Z=\prod_{k}Z_{k}$, $\omega_{k}$ is the frequency of the $k$-th
bosonic mode, and $m_{k}$ is the number of bosons in the $k$-th
mode. \cite{schlosshauer book,huang book,mandel book}
\end{singlespace}
\begin{singlespace}

\subsection{Equation of motion}
\end{singlespace}

The first-order effective Hamiltonian in the equation of motion (see
Appendix A for calculation details) is found to vanish, 
\begin{equation}
H_{eff}^{I}(t)=0,\label{eq:L1=00003D0}
\end{equation}
which means the system-bath interaction does not have first-order
contribution to the TLS's reduced dynamics in this case.

\begin{singlespace}
Introducing the following definitions with $\omega_{k0}\equiv\omega_{k}-\omega_{0}$
for notational convenience,
\begin{eqnarray}
D_{R}(t) & \equiv & \int_{0}^{t}dt'\sum_{k}|g_{k}|^{2}\bar{N_{k}}\cos\left(\omega_{k0}(t-t')\right),\label{eq:Dr def}\\
D_{I}(t) & \equiv & \int_{0}^{t}dt'\sum_{k}|g_{k}|^{2}\bar{N_{k}}\sin\left(\omega_{k0}(t-t')\right),\label{eq:Di def}\\
D'_{R}(t) & \equiv & \int_{0}^{t}dt'\sum_{k}|g_{k}|^{2}\left(\bar{N_{k}}+1\right)\cos\left(\omega_{k0}(t-t')\right),\label{eq:Dr' def}\\
D'_{I}(t) & \equiv & \int_{0}^{t}dt'\sum_{k}|g_{k}|^{2}\left(\bar{N_{k}}+1\right)\sin\left(\omega_{k0}(t-t')\right),\label{eq:Di' def}
\end{eqnarray}
where we have denoted the average occupation number in the $k$-th
mode of the bath as 
\begin{equation}
\bar{N_{k}}\equiv Tr_{E}\left(\rho_{E0}b_{k}^{\dagger}b_{k}\right)=\frac{1}{Z_{k}}\sum_{m_{k}=0}^{\infty}e^{-m_{k}\beta\omega_{k}}\langle m_{k}|b_{k}^{\dagger}b_{k}|m_{k}\rangle,
\end{equation}
it can be shown that the second-order term in the equation of motion
is (see Appendix A for calculation details) 
\begin{eqnarray}
\mathcal{L}_{2,t}\left(\rho\right) & = & -i\,\left[H_{eff}^{II}(t),\,\rho\right]-D_{R}(t)\left(\sigma_{-}\sigma_{+}\rho+\rho\sigma_{-}\sigma_{+}-2\sigma_{+}\rho\sigma_{-}\right)\nonumber \\
 &  & -D'_{R}(t)\left(\sigma_{+}\sigma_{-}\rho+\rho\sigma_{+}\sigma_{-}-2\sigma_{-}\rho\sigma_{+}\right),\label{eq:L2 final}
\end{eqnarray}
where the second-order effective Hamiltonian is defined as 
\begin{equation}
H_{eff}^{II}(t)\equiv D_{I}(t)\sigma_{-}\sigma_{+}-D'_{I}(t)\sigma_{+}\sigma_{-}.\label{eq:H eff 2}
\end{equation}

With Eq.(\ref{eq:L1=00003D0}) for the first-order term and Eq.(\ref{eq:L2 final})
for the second-order term, we can write down the equation of motion
up to second order, 
\begin{eqnarray}
\frac{d}{dt}\rho_{t} & = & -i\,\left[H_{eff}^{II}(t),\,\rho_{t}\right]-D_{R}(t)\left(\sigma_{-}\sigma_{+}\rho_{t}+\rho_{t}\sigma_{-}\sigma_{+}-2\sigma_{+}\rho_{t}\sigma_{-}\right)\nonumber \\
 &  & -D'_{R}(t)\left(\sigma_{+}\sigma_{-}\rho_{t}+\rho_{t}\sigma_{+}\sigma_{-}-2\sigma_{-}\rho_{t}\sigma_{+}\right),\label{eq:eom 2nd order}
\end{eqnarray}
where the second-order effective Hamiltonian $H_{eff}^{II}(t)$ is
defined in Eq.(\ref{eq:H eff 2}) and the prefactors $D_{R}(t)$,
$D_{I}(t)$, $D'_{R}(t)$, and $D'_{I}(t)$ are defined in Eqs.(\ref{eq:Dr def},
\ref{eq:Di def}, \ref{eq:Dr' def}, \ref{eq:Di' def}) respectively.
Non-Markovian master equations like this are previously known in the
literature. For example, Eq.(5) of Ref.\cite{clos} shows a similar
master equation for a TLS, without the rotating wave approximation. 
\end{singlespace}
\begin{singlespace}

\subsubsection*{Decoherence rate}
\end{singlespace}

\begin{singlespace}
Loosely speaking, the prefactor $D_{R}(t)$ ($D'_{R}(t)$) may be
called ``decoherence rate'', \cite{schlosshauer book} which determines
how fast quantum coherence (as represented by some off-diagonal element
of the system's reduced density matrix in the relevant basis) decays.
By examining the formal expression of $D_{R}(t)$ ($D'_{R}(t)$) as
in Eq.(\ref{eq:Dr def}) (Eq.(\ref{eq:Dr' def})), 
\begin{eqnarray}
D_{R}(t) & = & \int_{0}^{t}dt'\sum_{k}|g_{k}|^{2}\bar{N_{k}}\cos\left(\omega_{k0}(t-t')\right),\\
D'_{R}(t) & = & \int_{0}^{t}dt'\sum_{k}|g_{k}|^{2}\bar{N_{k}}\cos\left(\omega_{k0}(t-t')\right)\nonumber \\
 &  & +\int_{0}^{t}dt'\sum_{k}|g_{k}|^{2}\cos\left(\omega_{k0}(t-t')\right),
\end{eqnarray}
we have the following observations: 

(a) For each occupied bosonic mode ($\bar{N_{k}}\neq0$), its contribution
to the decoherence rate depends linearly on its average occupation
number $\bar{N_{k}}$. This linear dependence on occuptation number
is well known. See also Eq.(3.219) of Ref.\cite{breuer book} for
another example of linear dependence on occupation number (albeit
at the transition frequency, in the case of a Markovian master equation). 

(b.1) For each occupied bosonic mode ($\bar{N_{k}}\neq0$), its contribution
to the decoherence rate is quadratic on its coupling strength to the
system $|g_{k}|$; and (b.2) in addition to the contributions from
occupied modes as discussed in (a) and (b.1), all modes coupled to
the system ($g_{k}\neq0$), regardless of being occupied or unoccupied,
contribute to the prefactor $D'_{R}(t)$ for the last term in Eq.(\ref{eq:eom 2nd order}),
and each coupled mode's contribution is quadratic on its coupling
strength to the system $|g_{k}|$. This quadratic dependence on coupling
strength is also well known in the literature. For example, the second-order
contribution in Eqs.(16, 33) of Ref.\cite{vacchini} shows another
example of quadratic dependence, though with the environment initially
in the vacuum state. 
\end{singlespace}
\begin{singlespace}

\subsubsection*{Constant decoherence rate}
\end{singlespace}

\begin{singlespace}
Generally, decoherence rates $D_{R}(t)$ ($D'_{R}(t)$) can depend
on time. In many cases, however, decoherence rates are (approximately)
time independent. Appendix B shows one way constant decoherence rates
can be recovered. \footnote{Note that the discussions in Appendix B regarding the evaluation of
prefactors like $D_{R}(t)$ are not necessarily rigorous and are meant
for heuristic purpose. We follow the treatments and arguments as in
references \cite{heitler book,jones,heller book}, which are supposedly
standard practice but are not necessarily always valid. Figures 1-3
are for illustrative purpose and are by no means accurate.} (Also note that Markovian master equations usually come with constant
decoherence rates, which are extensively studied in the literature.
See, for example, Eq.(3.219) of Ref.\cite{breuer book} for a Markovian
equation for a TLS.) A constant decoherence rate in turn implies exponential
decay in relevant elements of the system's reduced density matrix
$\rho_{t}$. 
\end{singlespace}
\begin{singlespace}

\subsubsection*{Vacuum limit}
\end{singlespace}

\begin{singlespace}
Suppose the bosonic field is initially in the vacuum state, $\rho_{E0}=|0\rangle\langle0|$.
\footnote{The vacuum state may be throught of as the ``thermal state'' at
zero temperature. Formally, the vacuum state is diagonal in the occupation
number eigenbasis, therefore the derivations leading to Eq.(\ref{eq:eom 2nd order})
remains valid.} This specific case of a TLS coupled to a bath initially in the vacuum
state is previously studied in \cite{vacchini}. In this vacuum limit,
the expected occupation number is zero for all bosonic field modes,
\begin{equation}
\bar{N_{k}}\equiv Tr_{E}\left(\rho_{E0}b_{k}^{\dagger}b_{k}\right)=0.\label{eq:Nk=00003D0}
\end{equation}
Plugging Eq.(\ref{eq:Nk=00003D0}) into Eqs.(\ref{eq:Dr def}-\ref{eq:Di' def}),
we have 
\begin{eqnarray}
D_{R}(t) & = & 0,\label{eq:Dr vanishes}\\
D_{I}(t) & = & 0,\label{eq:Di vanishes}\\
D'_{R}(t) & = & \int_{0}^{t}dt'\sum_{k}|g_{k}|^{2}\cos\left(\omega_{k0}(t-t')\right)\nonumber \\
 & = & Re\left(\int_{0}^{t}dt'\sum_{k}|g_{k}|^{2}e^{-i\omega_{k0}(t-t')}\right)\equiv\frac{1}{2}\gamma^{(2)}(t),\label{eq:gamma}\\
D'_{I}(t) & = & \int_{0}^{t}dt'\sum_{k}|g_{k}|^{2}\sin\left(\omega_{k0}(t-t')\right)\nonumber \\
 & = & -Im\left(\int_{0}^{t}dt'\sum_{k}|g_{k}|^{2}e^{-i\omega_{k0}(t-t')}\right)\equiv-\frac{1}{2}S^{(2)}(t),\label{eq:S}
\end{eqnarray}
where new parameters $\gamma^{(2)}(t)$ and $S^{(2)}(t)$ have been
introduced in accordance with the notations in Eqs.(33, 16) of Ref.\cite{vacchini}. 

Plugging Eqs.(\ref{eq:Dr vanishes}-\ref{eq:S}) into Eqs.(\ref{eq:H eff 2},
\ref{eq:eom 2nd order}), we obtain the equation of motion describing
the reduced dynamics of a TLS coupled to a bosonic field initially
in the vacuum state (up to second order): 
\begin{eqnarray}
\frac{d}{dt}\rho_{t} & = & -i\,\left[\frac{1}{2}S^{(2)}(t)\sigma_{+}\sigma_{-},\,\rho_{t}\right]-\frac{1}{2}\gamma^{(2)}(t)\left(\sigma_{+}\sigma_{-}\rho_{t}+\rho_{t}\sigma_{+}\sigma_{-}-2\sigma_{-}\rho_{t}\sigma_{+}\right)\nonumber \\
 & = & -\frac{i}{2}S^{(2)}(t)\,\left[\sigma_{+}\sigma_{-},\,\rho_{t}\right]+\gamma^{(2)}(t)\left(\sigma_{-}\rho_{t}\sigma_{+}-\frac{1}{2}\left\{ \sigma_{+}\sigma_{-},\,\rho_{t}\right\} \right).\label{eq:eom vacuum}
\end{eqnarray}
Comparing Eq.(\ref{eq:eom vacuum}) with Eqs.(26, 28, 33, 16) of Ref.\cite{vacchini},
we see that our result agrees with the second-order result in \cite{vacchini}. 
\end{singlespace}
\begin{singlespace}

\subsection{Reduced dynamics}
\end{singlespace}

\begin{singlespace}
Now we use the second-order master equation Eq.(\ref{eq:eom 2nd order})
to easily get some quantitative results and gain more insights into
the TLS coupled to bosonic field.
\end{singlespace}
\begin{singlespace}

\subsubsection*{Differential equations for density matrix elements}
\end{singlespace}

\begin{singlespace}
To find the equations of motion for the elements $\rho_{mn}(t)$ of
the reduced density matrix $\rho_{t}$, we sandwich both sides of
Eq.(\ref{eq:eom 2nd order}) with $\langle m|\ldots|n\rangle$ for
$m,\,n=0,\,1$, with the convention that $|0\rangle$ represents spin-up
and $|1\rangle$ represents spin-down. With $\sigma_{+}|0\rangle=0$,
$\sigma_{+}|1\rangle=2|0\rangle$, $\sigma_{-}|0\rangle=2|1\rangle$,
and $\sigma_{-}|1\rangle=0$, it can be shown that the evolution of
matrix elements are governed by a system of linear ordinary differential
equations as follows, 
\begin{eqnarray}
\frac{d}{dt}\rho_{00}(t) & = & -8D'_{R}(t)\rho_{00}(t)+8D{}_{R}(t)\rho_{11}(t),\label{eq:rho 00}\\
\frac{d}{dt}\rho_{01}(t) & = & i\left(4\left(D_{I}(t)+D'_{I}(t)\right)\right)\rho_{01}(t)-4\left(D_{R}(t)+D'_{R}(t)\right)\rho_{01}(t),\label{eq:rho 01}\\
\frac{d}{dt}\rho_{10}(t) & = & -i\left(4\left(D_{I}(t)+D'_{I}(t)\right)\right)\rho_{10}(t)-4\left(D_{R}(t)+D'_{R}(t)\right)\rho_{10}(t),\label{eq:rho 10}\\
\frac{d}{dt}\rho_{11}(t) & = & 8D'_{R}(t)\rho_{00}(t)-8D{}_{R}(t)\rho_{11}(t),\label{eq:rho 11}
\end{eqnarray}

We see that the evolution of off-diagonal element $\rho_{01}(t)$
is governed by a (linear homogeneous) ordinary differential equation
Eq.(\ref{eq:rho 01}), that is, the dynamics of $\rho_{01}(t)$ is
decoupled from that of the other density matrix elements. The same
can be said about $\rho_{10}(t)$. For the diagonal elements $\rho_{00}(t)$
and $\rho_{11}(t)$, they form a system of coupled differential equations.
\end{singlespace}
\begin{singlespace}

\subsubsection*{General solutions for coherence}
\end{singlespace}

\begin{singlespace}
We can solve the homogeneous linear ODE for the off-diagonals $\rho_{01}(t)$
and $\rho_{10}(t)$, \cite{birkhoff book}
\begin{eqnarray}
\rho_{01}(t) & = & \rho_{01}(0)\exp\left(i\int_{0}^{t}dt'\,4\left(D_{I}(t')+D'_{I}(t')\right)\right)\nonumber \\
 &  & \exp\left(-\int_{0}^{t}dt'\,4\left(D_{R}(t')+D'_{R}(t')\right)\right),\label{eq:rho 01 solution}\\
\rho_{10}(t) & = & \rho_{10}(0)\exp\left(-i\int_{0}^{t}dt'\,4\left(D_{I}(t')+D'_{I}(t')\right)\right)\nonumber \\
 &  & \exp\left(-\int_{0}^{t}dt'\,4\left(D_{R}(t')+D'_{R}(t')\right)\right).\label{eq:rho 10 solution}
\end{eqnarray}

As we can see, the first term in Eq.(\ref{eq:rho 01}) with a pure
imaginary prefactor results in a phase shift of $\rho_{01}(t)$, as
is manifest in the first exponential factor of the solution Eq.(\ref{eq:rho 01 solution});
the second term in Eq.(\ref{eq:rho 01}) with a real prefactor results
in a decay in the amplitude of $\rho_{01}(t)$, as is manifest in
the second exponential factor of Eq.(\ref{eq:rho 01 solution}). The
same can be said about $\rho_{10}(t)$. Focusing on the amplitude
of $\rho_{01}(t)$ ($\rho_{10}(t)$), we see that 
\begin{eqnarray}
|\rho_{01}(t)| & = & |\rho_{01}(0)|\exp\left(-\int_{0}^{t}dt'\,4\left(D_{R}(t')+D'_{R}(t')\right)\right),\\
|\rho_{10}(t)| & = & |\rho_{10}(0)|\exp\left(-\int_{0}^{t}dt'\,4\left(D_{R}(t')+D'_{R}(t')\right)\right).
\end{eqnarray}
Thus we see that the coherence $\rho_{01}(t)$ ($\rho_{10}(t)$) between
the system's energy eigenlevels decay in this case. 
\end{singlespace}
\begin{singlespace}

\subsubsection*{General solutions for populations}
\end{singlespace}

\begin{singlespace}
To solve for the diagonals $\rho_{00}(t)$ and $\rho_{11}(t)$, that
is, the spin-up and spin-down populations, we may make use of the
unit trace property of density matrix, namely $\rho_{00}(t)+\rho_{11}(t)=1$.
Plugging $\rho_{11}(t)=1-\rho_{00}(t)$ into Eq.(\ref{eq:rho 00}),
we obtain a linear inhomogeneous ODE for $\rho_{00}(t)$, 
\begin{eqnarray}
\frac{d}{dt}\rho_{00}(t) & = & -8D'_{R}(t)\rho_{00}(t)+8D{}_{R}(t)\left(1-\rho_{00}(t)\right),\\
\Rightarrow\qquad\frac{d}{dt}\rho_{00}(t) & = & -8\left(D_{R}(t)+D'_{R}(t)\right)\rho_{00}(t)+8D_{R}(t),
\end{eqnarray}
the solution to which is \cite{birkhoff book}
\begin{eqnarray}
\rho_{00}(t) & = & \rho_{00}(0)\exp\left(-\int_{0}^{t}dt'\,8\left(D_{R}(t')+D'_{R}(t')\right)\right)\nonumber \\
 &  & +\exp\left(-\int_{0}^{t}dt'\,8\left(D_{R}(t')+D'_{R}(t')\right)\right)\nonumber \\
 &  & \times\int_{0}^{t}dt'\,8D{}_{R}(t')\exp\left(\int_{0}^{t'}dt"\,8\left(D_{R}(t")+D'_{R}(t")\right)\right).\label{eq:rho 00 solution}
\end{eqnarray}
The spin-down population may also be obtained accordingly, 
\begin{equation}
\rho_{11}(t)=1-\rho_{00}(t).\label{eq:rho 11 solution}
\end{equation}
 
\end{singlespace}
\begin{singlespace}

\subsubsection*{High temperature limit}
\end{singlespace}

\begin{singlespace}
If the bath starts at (extremely) high temperature, the average number
of bosons in the field modes are large, \cite{mandel book} that is,
$\bar{N_{k}}\equiv Tr_{E}\left(\rho_{E0}b_{k}^{\dagger}b_{k}\right)\gg1$,
in which case we can treat $\bar{N_{k}}+1\cong\bar{N_{k}}$ in Eq.(\ref{eq:Dr' def})
for $D'_{R}(t)$, 
\begin{eqnarray}
D'_{R}(t) & \equiv & \int_{0}^{t}dt'\sum_{k}|g_{k}|^{2}\left(\bar{N_{k}}+1\right)\cos\left(\omega_{k0}(t-t')\right)\nonumber \\
 & \cong & \int_{0}^{t}dt'\sum_{k}|g_{k}|^{2}\bar{N_{k}}\cos\left(\omega_{k0}(t-t')\right)\nonumber \\
 & = & D_{R}(t),\label{eq:Dr'=00003DDr}
\end{eqnarray}
and in Eq.(\ref{eq:Di' def}) for $D'_{I}(t)$, 
\begin{eqnarray}
D'_{I}(t) & \equiv & \int_{0}^{t}dt'\sum_{k}|g_{k}|^{2}\left(\bar{N_{k}}+1\right)\sin\left(\omega_{k0}(t-t')\right)\nonumber \\
 & \cong & \int_{0}^{t}dt'\sum_{k}|g_{k}|^{2}\bar{N_{k}}\sin\left(\omega_{k0}(t-t')\right)\nonumber \\
 & = & D_{I}(t).\label{eq:Di'=00003DDi}
\end{eqnarray}

Plugging Eqs.(\ref{eq:Dr'=00003DDr}, \ref{eq:Di'=00003DDi}) into
Eqs.(\ref{eq:rho 01 solution}, \ref{eq:rho 10 solution}), we see
that the coherence between energy eigenlevels will evolve as 
\begin{eqnarray}
\rho_{01}(t) & = & \rho_{01}(0)\exp\left(i\,8\int_{0}^{t}dt'D_{I}(t')\right)\exp\left(-8\int_{0}^{t}dt'D_{R}(t')\right),\\
\rho_{10}(t) & = & \rho_{10}(0)\exp\left(-i\,8\int_{0}^{t}dt'D_{I}(t')\right)\exp\left(-8\int_{0}^{t}dt'D_{R}(t')\right),
\end{eqnarray}
with the amplitudes decaying according to 
\begin{eqnarray}
|\rho_{01}(t)| & = & |\rho_{01}(0)|\exp\left(-8\int_{0}^{t}dt'D_{R}(t')\right),\\
|\rho_{10}(t)| & = & |\rho_{10}(0)|\exp\left(-8\int_{0}^{t}dt'D_{R}(t')\right).
\end{eqnarray}

Plugging Eqs.(\ref{eq:Dr'=00003DDr}, \ref{eq:Di'=00003DDi}) into
Eq.(\ref{eq:rho 00 solution}), we see that the spin-up population
evolves as 
\begin{eqnarray}
\rho_{00}(t) & = & \rho_{00}(0)\exp\left(-16\int_{0}^{t}dt'D_{R}(t')\right)\nonumber \\
 &  & +\frac{1}{2}\exp\left(-16\int_{0}^{t}dt'D_{R}(t')\right)\int_{0}^{t}dt'\,16D{}_{R}(t')\exp\left(16\int_{0}^{t'}dt"D_{R}(t")\right)\nonumber \\
 & = & \rho_{00}(0)\exp\left(-16\int_{0}^{t}dt'D_{R}(t')\right)\nonumber \\
 &  & +\frac{1}{2}\exp\left(-16\int_{0}^{t}dt'D_{R}(t')\right)\left(\exp\left(16\int_{0}^{t'}dt"D_{R}(t")\right)|_{t'=0}^{t'=t}\right)\nonumber \\
 & = & \rho_{00}(0)\exp\left(-16\int_{0}^{t}dt'D_{R}(t')\right)+\frac{1}{2}\left(1-\exp\left(-16\int_{0}^{t}dt'D_{R}(t')\right)\right)\nonumber \\
 & = & \frac{1}{2}+\left(\rho_{00}(0)-\frac{1}{2}\right)\exp\left(-16\int_{0}^{t}dt'D_{R}(t')\right).\label{eq:rho 00 high T}
\end{eqnarray}

From Eq.(\ref{eq:rho 00 high T}), we may make two observations about
the population at the high temperature limit: 

(a) If we start at $\rho_{00}(0)=\frac{1}{2}$, it will stay at $\rho_{00}(t)=\frac{1}{2}$
subsequently. In other words, $\rho_{00}(t)=\frac{1}{2}$ is a steady
state solution.

(b) Regardless of the initial spin-up population, even for $\rho_{00}(0)\neq\frac{1}{2}$,
as long as sufficient time passes by so that the factor $\exp\left(-16\int_{0}^{t}dt'D_{R}(t')\right)$
gets close enough to vanishing,\footnote{Suppose that it is within the domain of applicability of our master
equation formalism, namely reasonably short time and/or weak coupling,
and that the second-order approximate equation of motion still holds.} we may say the spin-up population approaches the steady state solution
$\rho_{00}=\frac{1}{2}$. By Eq.(\ref{eq:rho 11 solution}), the spin-down
population will also be $\rho_{11}=1-\rho_{00}=\frac{1}{2}$ in this
case.

These observations are consistent with statistical mechanics - at
the high temperature limit, the energy eigenlevels should be equally
populated at equilibrium. \cite{pathria book}
\end{singlespace}
\begin{singlespace}

\subsubsection*{Low temperature limit}
\end{singlespace}

\begin{doublespace}
If the bath starts at zero temperature, where the average number of
bosons in the field modes are zero, \cite{mandel book} that is, $\bar{N_{k}}\equiv Tr_{E}\left(\rho_{E0}b_{k}^{\dagger}b_{k}\right)=0$,
the coefficients of the linear differential equations become 
\begin{eqnarray}
D_{R}(t) & \equiv & \int_{0}^{t}dt'\sum_{k}|g_{k}|^{2}\bar{N_{k}}\cos\left(\omega_{k0}(t-t')\right)=0,\label{eq:Dr low T}\\
D_{I}(t) & \equiv & \int_{0}^{t}dt'\sum_{k}|g_{k}|^{2}\bar{N_{k}}\sin\left(\omega_{k0}(t-t')\right)=0,\\
D'_{R}(t) & \equiv & \int_{0}^{t}dt'\sum_{k}|g_{k}|^{2}\left(\bar{N_{k}}+1\right)\cos\left(\omega_{k0}(t-t')\right)\nonumber \\
 & = & \int_{0}^{t}dt'\sum_{k}|g_{k}|^{2}\cos\left(\omega_{k0}(t-t')\right)\nonumber \\
 & \equiv & D_{R}^{0}(t),\\
D'_{I}(t) & \equiv & \int_{0}^{t}dt'\sum_{k}|g_{k}|^{2}\left(\bar{N_{k}}+1\right)\sin\left(\omega_{k0}(t-t')\right)\nonumber \\
 & = & \int_{0}^{t}dt'\sum_{k}|g_{k}|^{2}\sin\left(\omega_{k0}(t-t')\right)\nonumber \\
 & \equiv & D_{I}^{0}(t).\label{eq:Di' low T}
\end{eqnarray}

\end{doublespace}

\begin{singlespace}
Plugging Eqs.(\ref{eq:Dr low T}-\ref{eq:Di' low T}) into Eqs.(\ref{eq:rho 01 solution},
\ref{eq:rho 10 solution}), we see that the coherence will now evolve
as 
\begin{eqnarray}
\rho_{01}(t) & = & \rho_{01}(0)\exp\left(i\,4\int_{0}^{t}dt'D_{I}^{0}(t')\right)\exp\left(-4\int_{0}^{t}dt'D_{R}^{0}(t')\right),\\
\rho_{10}(t) & = & \rho_{10}(0)\exp\left(-i\,4\int_{0}^{t}dt'D_{I}^{0}(t')\right)\exp\left(-4\int_{0}^{t}dt'D_{R}^{0}(t')\right);
\end{eqnarray}
and their amplitudes decaying according to 
\begin{eqnarray}
|\rho_{01}(t)| & = & |\rho_{01}(0)|\exp\left(-4\int_{0}^{t}dt'D_{R}^{0}(t')\right),\\
|\rho_{10}(t)| & = & |\rho_{10}(0)|\exp\left(-4\int_{0}^{t}dt'D_{R}^{0}(t')\right).
\end{eqnarray}

Plugging Eqs.(\ref{eq:Dr low T}-\ref{eq:Di' low T}) into Eq.(\ref{eq:rho 00 solution}),
we see that the spin-up population now evolves as 
\begin{eqnarray}
\rho_{00}(t) & = & \rho_{00}(0)\exp\left(-8\int_{0}^{t}dt'D_{R}^{0}(t')\right).\label{eq:rho 00 low T}
\end{eqnarray}

From Eq.(\ref{eq:rho 00 low T}), we may make two observations about
the population at zero temperature: 

(a) If we start at $\rho_{00}(0)=0$, it will stay at $\rho_{00}(t)=0$.
In other words, $\rho_{00}(t)=0$ is a steady state solution, and
thus by Eq.(\ref{eq:rho 11 solution}) $\rho_{11}(t)=1-\rho_{00}(t)=1$,
that is, all populations being in spin-down (the energy ground state).

(b) Regardless of the initial spin-up population, even for $\rho_{00}(0)\neq0$,
as long as sufficient time passes by so that the factor $\exp\left(-8\int_{0}^{t}dt'D_{R}^{0}(t')\right)$
gets close enough to vanishing,\footnote{Suppose that it is within the domain of applicability of our master
equation formalism, namely reasonably short time and/or weak coupling,
and that the second-order approximate equation of motion still holds.} we may say the spin-up population approaches the steady state solution
$\rho_{00}=0$, which also implies $\rho_{11}=1-\rho_{00}=1$ by Eq.(\ref{eq:rho 11 solution}).

These observations are consistent with statistical mechanics - at
zero temperature, the equilibrium population should be all in the
ground state. \cite{pathria book}
\end{singlespace}

\section{Conclusions}

We develop a formally exact master equation for open quantum systems
in a succint and intuitive way. Our derivation is direct and \textquotedblleft by
construction\textquotedblright . In particular, it dispenses with
the use of an \textquotedblleft inverse\textquotedblright{} map, which
was used by previous derivations of formally exact master equations.
Applying our formalism to the spin-boson model at arbitrary temperature,
we observe non-exponential decoherence and relaxation characteristic
of non-Markovian behaviors. The equation of motion obtained herein,
albeit a second-order approximation, yields the right steady state
solution, in agreement with standard statistical mechanical predictions.
The formalism can be applied to study more physical examples and further
explore its usefulness. For example, it can be used to study the dynamics
of two atoms in an optical cavity, which could have implications on
two-atom entanglement \cite{reiter}. Higher-order equations of motion
can also be obtained mechanically using Eqs.(\ref{eq:1. eom higher order big O}-\ref{eq:1. eom higher order})
to study corrections to second-order dynamics. 

\subsection*{Appendix A}

\begin{singlespace}
To derive the equation of motion for the TLS's reduced density matrix,
we first caste the full interaction Hamiltonian Eq.(\ref{eq:int-H full})
into the form of Eq.(\ref{eq:1. product form}):
\begin{equation}
H_{SE}(t)=S_{1}\otimes E_{1}(t)+S_{2}\otimes E_{2}(t),\label{eq:int-H re-written}
\end{equation}
where the system operators are defined as 
\begin{eqnarray}
S_{1} & \equiv & \sigma_{+},\label{eq:S1}\\
S_{2} & \equiv & \sigma_{-},\label{eq:S2}
\end{eqnarray}
and we have absorbed the time dependence into the bath operators,
\begin{eqnarray}
E_{1}(t) & \equiv & \sum_{k}g_{k}e^{-i(\omega_{k}-\omega_{0})t}b_{k},\label{eq:E1}\\
E_{2}(t) & \equiv & \sum_{k}g_{k}e^{i(\omega_{k}-\omega_{0})t}b_{k}^{\dagger}.\label{eq:E2}
\end{eqnarray}
Hereafter we shall denote $\omega_{k0}\equiv\omega_{k}-\omega_{0}$
for convenience.
\end{singlespace}
\begin{singlespace}

\subsubsection*{First-order term in the equation of motion}
\end{singlespace}

\begin{singlespace}
To evaluate the first-order term of the equation of motion, plugging
Eqs.(\ref{eq:S1}-\ref{eq:E2}) into Eq.(\ref{eq:1. L1 eom}) yields
\begin{eqnarray}
H_{eff}^{I}(t) & \equiv & Tr_{E}\left(\rho_{E0}E_{1}(t)\right)S_{1}+Tr_{E}\left(\rho_{E0}E_{2}(t)\right)S_{2}\nonumber \\
 & = & \sum_{k}g_{k}e^{-i\omega_{k0}t}Tr_{E}\left(\rho_{E0}b_{k}\right)S_{1}\nonumber \\
 &  & +\sum_{k}g_{k}e^{i\omega_{k0}t}Tr_{E}\left(\rho_{E0}b_{k}^{\dagger}\right)S_{2}.
\end{eqnarray}
The prefactor $Tr_{E}\left(\rho_{E0}b_{k}\right)$ for an arbitray
$k$-th mode can be evaluated as 
\begin{eqnarray}
Tr_{E}\left(\rho_{E0}b_{k}\right) & = & Tr_{Ek}\left(\frac{1}{Z_{k}}\sum_{m_{k}=0}^{\infty}e^{-m_{k}\beta\omega_{k}}|m_{k}\rangle\langle m_{k}|b_{k}\right)\nonumber \\
 &  & \prod_{k'\neq k}Tr_{Ek'}\left(\frac{1}{Z_{k'}}\sum_{m_{k'}=0}^{\infty}e^{-m_{k'}\beta\omega_{k'}}|m_{k'}\rangle\langle m_{k'}|\right)\nonumber \\
 & = & \frac{1}{Z_{k}}\sum_{m_{k}=0}^{\infty}e^{-m_{k}\beta\omega_{k}}\langle m_{k}|b_{k}|m_{k}\rangle\nonumber \\
 & = & 0,\label{eq:b=00003D0}
\end{eqnarray}
where $Tr_{Ek}\left(\ldots\right)$ denotes the partial trace over
the $k$-th bosonic mode. Similarly, the prefactor $Tr_{E}\left(\rho_{E0}b_{k}^{\dagger}\right)$
for an arbitray $k$-th mode is
\begin{eqnarray}
Tr_{E}\left(\rho_{E0}b_{k}^{\dagger}\right) & = & Tr_{Ek}\left(\frac{1}{Z_{k}}\sum_{m_{k}=0}^{\infty}e^{-m_{k}\beta\omega_{k}}|m_{k}\rangle\langle m_{k}|b_{k}^{\dagger}\right)\nonumber \\
 & = & \frac{1}{Z_{k}}\sum_{m_{k}=0}^{\infty}e^{-m_{k}\beta\omega_{k}}\langle m_{k}|b_{k}^{\dagger}|m_{k}\rangle\nonumber \\
 & = & 0.\label{eq:b^=00003D0}
\end{eqnarray}
Thus the first-order effective Hamiltonian vanishes, 
\begin{equation}
H_{eff}^{I}(t)=0,\label{eq:L1=00003D0-1}
\end{equation}
which means the system-bath interaction does not have first-order
contribution to the TLS's reduced dynamics in this case.
\end{singlespace}
\begin{singlespace}

\subsubsection*{Second-order term in the equation of motion}
\end{singlespace}

\begin{singlespace}
To evaluate the second-order term of the equation of motion, plugging
Eqs.(\ref{eq:S1}-\ref{eq:E2}) into Eq.(\ref{eq:1. L2 eom}) yields
\begin{equation}
\mathcal{L}_{2,t}\left(\rho\right)=-\sum_{m=1,2}\sum_{n=1,2}\int_{0}^{t}dt'\left(\mathcal{C}_{mn}(t,t')\left[S_{m},\:S_{n}\rho\right]-\mathcal{C}_{nm}(t',t)\left[S_{m},\:\rho S_{n}\right]\right),\label{eq:L2 step1}
\end{equation}
where the system operators $\left\{ S_{1},\,S_{2}\right\} $ are now
time-independent and the coefficients are defined as in Eq.(\ref{eq:1. L2 coeff}),
\begin{equation}
\mathcal{C}_{jk}(t,t')\equiv Tr_{E}\left(\rho_{E0}E_{j}(t)E_{k}(t')\right)-Tr_{E}\left(\rho_{E0}E_{j}(t)\right)Tr_{E}\left(\rho_{E0}E_{k}(t')\right).\label{eq:Cij old}
\end{equation}
To simplify Eq.(\ref{eq:Cij old}), we note that 
\begin{eqnarray}
Tr_{E}\left(\rho_{E0}E_{1}(t)\right) & = & \sum_{k}g_{k}e^{-i\omega_{k0}t}Tr_{E}\left(\rho_{E0}b_{k}\right)=0,\\
Tr_{E}\left(\rho_{E0}E_{2}(t)\right) & = & \sum_{k}g_{k}e^{i\omega_{k0}t}Tr_{E}\left(\rho_{E0}b_{k}^{\dagger}\right)=0,
\end{eqnarray}
where we have made use of Eqs.(\ref{eq:b=00003D0}, \ref{eq:b^=00003D0}).
Therefore, the coefficients are now 
\begin{equation}
\mathcal{C}_{jk}(t,t')=Tr_{E}\left(\rho_{E0}E_{j}(t)E_{k}(t')\right).\label{eq:Cij}
\end{equation}

Let's now evaluate Eq.(\ref{eq:L2 step1}) term by term.

For the term with $m=n=1$, the first coefficient is 
\begin{eqnarray}
\mathcal{C}_{11}(t,t') & = & Tr_{E}\left(\rho_{E0}E_{1}(t)E_{1}(t')\right)\nonumber \\
 & = & \sum_{k}\sum_{k'}g_{k}g_{k'}e^{-i\omega_{k0}t}e^{-i\omega_{k'0}t'}Tr_{E}\left(\rho_{E0}b_{k}b_{k'}\right).
\end{eqnarray}
We will show that the last factor $Tr_{E}\left(\rho_{E0}b_{k}b_{k'}\right)$
vanishes for arbitrary $\left(k,\,k'\right)$. First, for the case
of $k\neq k'$: 
\begin{eqnarray}
Tr_{E}\left(\rho_{E0}b_{k}b_{k'}\right) & = & Tr_{E}\left(\prod_{K}\left(\frac{1}{Z_{K}}\sum_{m_{K}=0}^{\infty}e^{-m_{K}\beta\omega_{K}}|m_{K}\rangle\langle m_{K}|\right)b_{k}b_{k'}\right)\nonumber \\
 & = & \frac{1}{Z_{k}Z_{k'}}Tr_{E_{k}}\left(\sum_{m_{k}=0}^{\infty}e^{-m_{k}\beta\omega_{k}}|m_{k}\rangle\langle m_{k}|b_{k}\right)\nonumber \\
 &  & Tr_{E_{k'}}\left(\sum_{m_{k'}=0}^{\infty}e^{-m_{k'}\beta\omega_{k'}}|m_{k'}\rangle\langle m_{k'}|b_{k'}\right)\nonumber \\
 & = & \frac{1}{Z_{k}Z_{k'}}\left(\sum_{m_{k}=0}^{\infty}e^{-m_{k}\beta\omega_{k}}\langle m_{k}|b_{k}|m_{k}\rangle\right)\nonumber \\
 &  & \left(\sum_{m_{k'}=0}^{\infty}e^{-m_{k'}\beta\omega_{k'}}\langle m_{k'}|b_{k'}|m_{k'}\rangle\right)\nonumber \\
 & = & 0,
\end{eqnarray}
because $\langle m_{k}|b_{k}|m_{k}\rangle\propto\langle m_{k}+1|m_{k}\rangle=0$
vanishes for an arbitrary $k$; second, for the case of $k=k'$: 
\begin{eqnarray}
Tr_{E}\left(\rho_{E0}b_{k}b_{k}\right) & = & \frac{1}{Z_{k}}Tr_{E_{k}}\left(\sum_{m_{k}=0}^{\infty}e^{-m_{k}\beta\omega_{k}}|m_{k}\rangle\langle m_{k}|b_{k}b_{k}\right)\nonumber \\
 & = & \frac{1}{Z_{k}}\left(\sum_{m_{k}=0}^{\infty}e^{-m_{k}\beta\omega_{k}}\langle m_{k}|b_{k}b_{k}|m_{k}\rangle\right)\nonumber \\
 & = & 0,
\end{eqnarray}
because $\langle m_{k}|b_{k}b_{k}|m_{k}\rangle\propto\langle m_{k}+2|m_{k}\rangle=0$
vanishes for an arbitrary $k$. Therefore, we have shown 
\begin{equation}
\mathcal{C}_{11}(t,t')=0.
\end{equation}
Similarly, the second coefficient is 
\begin{eqnarray}
\mathcal{C}_{11}(t',t) & = & Tr_{E}\left(\rho_{E0}E_{1}(t')E_{1}(t)\right)\nonumber \\
 & = & \sum_{k}\sum_{k'}g_{k}g_{k'}e^{-i\omega_{k0}t'}e^{-i\omega_{k'0}t}Tr_{E}\left(\rho_{E0}b_{k}b_{k'}\right)\nonumber \\
 & = & 0.
\end{eqnarray}
Therefore, the term for $m=n=1$ in Eq.(\ref{eq:L2 step1}) vanishes.

Similarly, for the term with $m=n=2$, the first coefficient is 
\begin{eqnarray}
\mathcal{C}_{22}(t,t') & = & Tr_{E}\left(\rho_{E0}E_{2}(t)E_{2}(t')\right)\nonumber \\
 & = & \sum_{k}\sum_{k'}g_{k}g_{k'}e^{i\omega_{k0}t}e^{i\omega_{k'0}t'}Tr_{E}\left(\rho_{E0}b_{k}^{\dagger}b_{k'}^{\dagger}\right)\nonumber \\
 & = & 0,
\end{eqnarray}
as it can be similarly shown that $Tr_{E}\left(\rho_{E0}b_{k}^{\dagger}b_{k'}^{\dagger}\right)=0$
for arbitrary $\left(k,\,k'\right)$. Likewise, the second coefficient
can be shown to vanish, $\mathcal{C}_{22}(t',t)=0$. Therefore, the
term for $m=n=2$ in Eq.(\ref{eq:L2 step1}) vanishes.

Thus we are left with the cross terms with $\left(m=1,\,n=2\right)$
and $\left(m=2,\,n=1\right)$ in Eq.(\ref{eq:L2 step1}): 
\begin{eqnarray}
\mathcal{L}_{2,t}\left(\rho\right) & = & -\int_{0}^{t}dt'\{\:\mathcal{C}_{12}(t,t')\left[S_{1},\:S_{2}\rho\right]-\mathcal{C}_{21}(t',t)\left[S_{1},\:\rho S_{2}\right]\nonumber \\
 &  & +\mathcal{C}_{21}(t,t')\left[S_{2},\:S_{1}\rho\right]-\mathcal{C}_{12}(t',t)\left[S_{2},\:\rho S_{1}\right]\:\}\nonumber \\
 & = & -\int_{0}^{t}dt'\{\,\mathcal{C}_{12}(t,t')\left(\sigma_{+}\sigma_{-}\rho-\sigma_{-}\rho\sigma_{+}\right)+\mathcal{C}_{12}(t',t)\left(\rho\sigma_{+}\sigma_{-}-\sigma_{-}\rho\sigma_{+}\right)\nonumber \\
 &  & +\mathcal{C}_{21}(t,t')\left(\sigma_{-}\sigma_{+}\rho-\sigma_{+}\rho\sigma_{-}\right)+\mathcal{C}_{21}(t',t)\left(\rho\sigma_{-}\sigma_{+}-\sigma_{+}\rho\sigma_{-}\right)\,\},\label{eq:L2 step2}
\end{eqnarray}
where in the second equality we have rearranged the order of the terms.
The prefactor of each term in Eq.(\ref{eq:L2 step2}) will be evaluated
as follows.

For the first term, 
\begin{eqnarray}
\mathcal{C}_{12}(t,t') & = & Tr_{E}\left(\rho_{E0}E_{1}(t)E_{2}(t')\right)\nonumber \\
 & = & \sum_{k}\sum_{k'}g_{k}g_{k'}e^{-i\omega_{k0}t}e^{i\omega_{k'0}t'}Tr_{E}\left(\rho_{E0}b_{k}b_{k'}^{\dagger}\right),\label{eq:C12 step1}
\end{eqnarray}
where the factor $Tr_{E}\left(\rho_{E0}b_{k}b_{k'}^{\dagger}\right)$
is, for $k\neq k'$:
\begin{eqnarray}
Tr_{E}\left(\rho_{E0}b_{k}b_{k'}^{\dagger}\right) & = & \frac{1}{Z_{k}Z_{k'}}Tr_{E_{k}}\left(\sum_{m_{k}=0}^{\infty}e^{-m_{k}\beta\omega_{k}}|m_{k}\rangle\langle m_{k}|b_{k}\right)\nonumber \\
 &  & Tr_{E_{k'}}\left(\sum_{m_{k'}=0}^{\infty}e^{-m_{k'}\beta\omega_{k'}}|m_{k'}\rangle\langle m_{k'}|b_{k'}^{\dagger}\right)\nonumber \\
 & = & \frac{1}{Z_{k}Z_{k'}}\left(\sum_{m_{k}=0}^{\infty}e^{-m_{k}\beta\omega_{k}}\langle m_{k}|b_{k}|m_{k}\rangle\right)\nonumber \\
 &  & \left(\sum_{m_{k'}=0}^{\infty}e^{-m_{k'}\beta\omega_{k'}}\langle m_{k'}|b_{k'}^{\dagger}|m_{k'}\rangle\right)\nonumber \\
 & = & 0,\label{eq:bb'^}
\end{eqnarray}
and for $k=k'$: 
\begin{eqnarray}
Tr_{E}\left(\rho_{E0}b_{k}b_{k}^{\dagger}\right) & = & \frac{1}{Z_{k}}Tr_{E_{k}}\left(\sum_{m_{k}=0}^{\infty}e^{-m_{k}\beta\omega_{k}}|m_{k}\rangle\langle m_{k}|b_{k}b_{k}^{\dagger}\right)\nonumber \\
 & = & \frac{1}{Z_{k}}\left(\sum_{m_{k}=0}^{\infty}e^{-m_{k}\beta\omega_{k}}\langle m_{k}|b_{k}b_{k}^{\dagger}|m_{k}\rangle\right)\nonumber \\
 & = & \frac{1}{Z_{k}}\left(\sum_{m_{k}=0}^{\infty}e^{-m_{k}\beta\omega_{k}}\langle m_{k}|\left(b_{k}^{\dagger}b_{k}+\mathbb{I}\right)|m_{k}\rangle\right)\nonumber \\
 & = & \frac{1}{Z_{k}}\left(\sum_{m_{k}=0}^{\infty}e^{-m_{k}\beta\omega_{k}}\langle m_{k}|b_{k}^{\dagger}b_{k}|m_{k}\rangle+\sum_{m_{k}=0}^{\infty}e^{-m_{k}\beta\omega_{k}}\right)\nonumber \\
 & = & \bar{N_{k}}+1,\label{eq:bb^}
\end{eqnarray}
where we have denoted the average occupation number in the $k$-th
mode of the bath as 
\begin{equation}
\bar{N_{k}}\equiv Tr_{E}\left(\rho_{E0}b_{k}^{\dagger}b_{k}\right)=\frac{1}{Z_{k}}\sum_{m_{k}=0}^{\infty}e^{-m_{k}\beta\omega_{k}}\langle m_{k}|b_{k}^{\dagger}b_{k}|m_{k}\rangle;
\end{equation}
plugging Eqs.(\ref{eq:bb'^}, \ref{eq:bb^}) into Eq.(\ref{eq:C12 step1})
yields 
\begin{eqnarray}
\mathcal{C}_{12}(t,t') & = & \sum_{k}\sum_{k'}g_{k}g_{k'}e^{-i\omega_{k0}t}e^{i\omega_{k'0}t'}Tr_{E}\left(\rho_{E0}b_{k}b_{k'}^{\dagger}\right)\nonumber \\
 & = & \sum_{k}|g_{k}|^{2}e^{-i\omega_{k0}(t-t')}Tr_{E}\left(\rho_{E0}b_{k}b_{k}^{\dagger}\right)\nonumber \\
 & = & \sum_{k}|g_{k}|^{2}\left(\bar{N_{k}}+1\right)e^{-i\omega_{k0}(t-t')}\nonumber \\
 & = & \sum_{k}|g_{k}|^{2}\left(\bar{N_{k}}+1\right)\left(\cos\left(\omega_{k0}(t-t')\right)-i\sin\left(\omega_{k0}(t-t')\right)\right).
\end{eqnarray}

Similarly, for the second term, 
\begin{eqnarray}
\mathcal{C}_{12}(t',t) & = & Tr_{E}\left(\rho_{E0}E_{1}(t')E_{2}(t)\right)\nonumber \\
 & = & \sum_{k}\sum_{k'}g_{k}g_{k'}e^{-i\omega_{k0}t'}e^{i\omega_{k'0}t}Tr_{E}\left(\rho_{E0}b_{k}b_{k'}^{\dagger}\right),\nonumber \\
 & = & \sum_{k}|g_{k}|^{2}e^{i\omega_{k0}(t-t')}Tr_{E}\left(\rho_{E0}b_{k}b_{k}^{\dagger}\right)\nonumber \\
 & = & \sum_{k}|g_{k}|^{2}\left(\bar{N_{k}}+1\right)e^{i\omega_{k0}(t-t')}\nonumber \\
 & = & \sum_{k}|g_{k}|^{2}\left(\bar{N_{k}}+1\right)\left(\cos\left(\omega_{k0}(t-t')\right)+i\sin\left(\omega_{k0}(t-t')\right)\right),
\end{eqnarray}
where in the third and fourth equalities we have made use of Eqs.(\ref{eq:bb'^},
\ref{eq:bb^}).

For the third term, 
\begin{eqnarray}
\mathcal{C}_{21}(t,t') & = & Tr_{E}\left(E_{2}(t)E_{1}(t')\rho_{E0}\right)\nonumber \\
 & = & \sum_{k}\sum_{k'}g_{k}g_{k'}e^{i\omega_{k0}t}e^{-i\omega_{k'0}t'}Tr_{E}\left(\rho_{E0}b_{k}^{\dagger}b_{k'}\right),\label{eq:C21 step1}
\end{eqnarray}
where the factor $Tr_{E}\left(\rho_{E0}b_{k}^{\dagger}b_{k'}\right)$
is, for $k\neq k'$: 

\begin{eqnarray}
Tr_{E}\left(\rho_{E0}b_{k}^{\dagger}b_{k'}\right) & = & \frac{1}{Z_{k}Z_{k'}}Tr_{E_{k}}\left(\sum_{m_{k}=0}^{\infty}e^{-m_{k}\beta\omega_{k}}|m_{k}\rangle\langle m_{k}|b_{k}^{\dagger}\right)\nonumber \\
 &  & Tr_{E_{k'}}\left(\sum_{m_{k'}=0}^{\infty}e^{-m_{k'}\beta\omega_{k'}}|m_{k'}\rangle\langle m_{k'}|b_{k'}\right)\nonumber \\
 & = & \frac{1}{Z_{k}Z_{k'}}\left(\sum_{m_{k}=0}^{\infty}e^{-m_{k}\beta\omega_{k}}\langle m_{k}|b_{k}^{\dagger}|m_{k}\rangle\right)\nonumber \\
 &  & \left(\sum_{m_{k'}=0}^{\infty}e^{-m_{k'}\beta\omega_{k'}}\langle m_{k'}|b_{k'}|m_{k'}\rangle\right)\nonumber \\
 & = & 0,\label{eq:b^b'}
\end{eqnarray}
and for $k=k'$: 
\begin{eqnarray}
Tr_{E}\left(\rho_{E0}b_{k}^{\dagger}b_{k}\right) & = & \frac{1}{Z_{k}}Tr_{E_{k}}\left(\sum_{m_{k}=0}^{\infty}e^{-m_{k}\beta\omega_{k}}|m_{k}\rangle\langle m_{k}|b_{k}^{\dagger}b_{k}\right)\nonumber \\
 & = & \frac{1}{Z_{k}}\left(\sum_{m_{k}=0}^{\infty}e^{-m_{k}\beta\omega_{k}}\langle m_{k}|b_{k}^{\dagger}b_{k}|m_{k}\rangle\right)\nonumber \\
 & = & \bar{N_{k}};\label{eq:b^b}
\end{eqnarray}
plugging Eqs.(\ref{eq:b^b'}, \ref{eq:b^b}) into Eq.(\ref{eq:C21 step1})
yields 
\begin{eqnarray}
\mathcal{C}_{21}(t,t') & = & \sum_{k}\sum_{k'}g_{k}g_{k'}e^{i\omega_{k0}t}e^{-i\omega_{k'0}t'}Tr_{E}\left(\rho_{E0}b_{k}^{\dagger}b_{k'}\right)\nonumber \\
 & = & \sum_{k}|g_{k}|^{2}e^{i\omega_{k0}(t-t')}Tr_{E}\left(\rho_{E0}b_{k}^{\dagger}b_{k}\right)\nonumber \\
 & = & \sum_{k}|g_{k}|^{2}\bar{N_{k}}e^{i\omega_{k0}(t-t')}\nonumber \\
 & = & \sum_{k}|g_{k}|^{2}\bar{N_{k}}\left(\cos\left(\omega_{k0}(t-t')\right)+i\sin\left(\omega_{k0}(t-t')\right)\right).
\end{eqnarray}

Similarly, for the fourth term, 
\begin{eqnarray}
\mathcal{C}_{21}(t',t) & = & Tr_{E}\left(E_{2}(t')E_{1}(t)\rho_{E0}\right)\nonumber \\
 & = & \sum_{k}\sum_{k'}g_{k}g_{k'}e^{i\omega_{k0}t'}e^{-i\omega_{k'0}t}Tr_{E}\left(\rho_{E0}b_{k}^{\dagger}b_{k'}\right)\nonumber \\
 & = & \sum_{k}|g_{k}|^{2}e^{-i\omega_{k0}(t-t')}Tr_{E}\left(\rho_{E0}b_{k}^{\dagger}b_{k}\right)\nonumber \\
 & = & \sum_{k}|g_{k}|^{2}\bar{N_{k}}e^{-i\omega_{k0}(t-t')}\nonumber \\
 & = & \sum_{k}|g_{k}|^{2}\bar{N_{k}}\left(\cos\left(\omega_{k0}(t-t')\right)-i\sin\left(\omega_{k0}(t-t')\right)\right),
\end{eqnarray}
where in the third and fourth equalities we have made use of Eqs.(\ref{eq:b^b'},
\ref{eq:b^b}).

Now, for convenience, let's introduce the following notations: 
\begin{eqnarray}
D_{R}(t) & \equiv & \int_{0}^{t}dt'\sum_{k}|g_{k}|^{2}\bar{N_{k}}\cos\left(\omega_{k0}(t-t')\right),\label{eq:Dr def-1}\\
D_{I}(t) & \equiv & \int_{0}^{t}dt'\sum_{k}|g_{k}|^{2}\bar{N_{k}}\sin\left(\omega_{k0}(t-t')\right),\label{eq:Di def-1}\\
D'_{R}(t) & \equiv & \int_{0}^{t}dt'\sum_{k}|g_{k}|^{2}\left(\bar{N_{k}}+1\right)\cos\left(\omega_{k0}(t-t')\right),\label{eq:Dr' def-1}\\
D'_{I}(t) & \equiv & \int_{0}^{t}dt'\sum_{k}|g_{k}|^{2}\left(\bar{N_{k}}+1\right)\sin\left(\omega_{k0}(t-t')\right),\label{eq:Di' def-1}
\end{eqnarray}
with which the prefactors can be rewritten as 
\begin{eqnarray}
\int_{0}^{t}dt'\mathcal{C}_{12}(t,t') & = & D'_{R}(t)-iD'_{I}(t),\label{eq:C12 new}\\
\int_{0}^{t}dt'\mathcal{C}_{12}(t',t) & = & D'_{R}(t)+iD'_{I}(t),\\
\int_{0}^{t}dt'\mathcal{C}_{21}(t,t') & = & D_{R}(t)+iD{}_{I}(t),\\
\int_{0}^{t}dt'\mathcal{C}_{21}(t',t) & = & D_{R}(t)-iD{}_{I}(t).\label{eq:C21' new}
\end{eqnarray}
Plugging the prefactors Eqs.(\ref{eq:C12 new}-\ref{eq:C21' new})
into Eq.(\ref{eq:L2 step2}) and combining terms with like prefactors,
we have 
\begin{eqnarray}
\mathcal{L}_{2,t}\left(\rho\right) & = & -D_{R}(t)\left(\sigma_{-}\sigma_{+}\rho+\rho\sigma_{-}\sigma_{+}-2\sigma_{+}\rho\sigma_{-}\right)\nonumber \\
 &  & -D'_{R}(t)\left(\sigma_{+}\sigma_{-}\rho+\rho\sigma_{+}\sigma_{-}-2\sigma_{-}\rho\sigma_{+}\right)\nonumber \\
 &  & -i\,\left(D_{I}(t)\left[\sigma_{-}\sigma_{+},\,\rho\right]-D'_{I}(t)\left[\sigma_{+}\sigma_{-},\,\rho\right]\right).\label{eq:L2 step3}
\end{eqnarray}

We may put Eq.(\ref{eq:L2 step3}) into a compact form, 
\begin{eqnarray}
\mathcal{L}_{2,t}\left(\rho\right) & = & -i\,\left[H_{eff}^{II}(t),\,\rho\right]-D_{R}(t)\left(\sigma_{-}\sigma_{+}\rho+\rho\sigma_{-}\sigma_{+}-2\sigma_{+}\rho\sigma_{-}\right)\nonumber \\
 &  & -D'_{R}(t)\left(\sigma_{+}\sigma_{-}\rho+\rho\sigma_{+}\sigma_{-}-2\sigma_{-}\rho\sigma_{+}\right),\label{eq:L2 final-1}
\end{eqnarray}
where the second-order effective Hamiltonian is defined as 
\begin{equation}
H_{eff}^{II}(t)\equiv D_{I}(t)\sigma_{-}\sigma_{+}-D'_{I}(t)\sigma_{+}\sigma_{-}.\label{eq:H eff 2-1}
\end{equation}

\end{singlespace}

\subsection*{Appendix B}

\begin{singlespace}
We can further evaluate the prefactors in Eq.(\ref{eq:eom 2nd order}),
for example, 
\begin{eqnarray}
D_{R}(t) & \equiv & \int_{0}^{t}dt'\sum_{k}|g_{k}|^{2}\bar{N_{k}}\cos\left(\omega_{k0}(t-t')\right)\nonumber \\
 & = & \int_{0}^{t}dt'\sum_{k}|g_{k}|^{2}\bar{N_{k}}Re\left(e^{i\omega_{k0}\left(t-t'\right)}\right).\label{eq:Dr new}
\end{eqnarray}

\end{singlespace}
\begin{singlespace}

\subsubsection*{Integrand}
\end{singlespace}

\begin{singlespace}
First, let's examine the integrand in Eq.(\ref{eq:Dr new}) $\sum_{k}|g_{k}|^{2}\bar{N_{k}}\cos\left(\omega_{k0}(t-t')\right)=\sum_{k}|g_{k}|^{2}\bar{N_{k}}Re\left(e^{i\omega_{k0}\left(t-t'\right)}\right)$
as a function of $t'$, as shown in Figure 1. The integrand is peaked
around $t'=t$, loosely because of the following reasons. On the one
hand, at $t'=t$, the factor $\cos\left(\omega_{k0}(t-t)\right)=1$
for all $k$'s, therefore the sum $\sum_{k}|g_{k}|^{2}\bar{N_{k}}$
consists of positive terms $|g_{k}|^{2}\bar{N_{k}}$, all of which
add up constructively, leading to the peak at $t'=t$. On the other
hand, at $t'\neq t$, the factor $\cos\left(\omega_{k0}(t-t')\right)$
oscillates across various $k$'s, therefore contributions from various
terms with different $k$'s tend to cancel out each other. Loosely
speaking, the larger $|t-t'|$ is, the more oscillatory the factor
$\cos\left(\omega_{k0}(t-t')\right)$ becomes with respect to different
$k$'s, the more ``destructively'' the various terms $|g_{k}|^{2}\bar{N_{k}}\cos\left(\omega_{k0}(t-t')\right)$
interfere with one another, the smaller the sum $\qquad$ $\sum_{k}|g_{k}|^{2}\bar{N_{k}}\cos\left(\omega_{k0}(t-t')\right)$
becomes. This loosely explains the shape of the integrand $\sum_{k}|g_{k}|^{2}\bar{N_{k}}\cos\left(\omega_{k0}(t-t')\right)$
as a function of $t'$. (See \cite{jones} for similar discussions
regarding the peak of the integrand $\sum_{k}|g_{k}|^{2}\bar{N_{k}}Re\left(e^{i\omega_{k0}\left(t-t'\right)}\right)$.)
\end{singlespace}
\begin{singlespace}

\subsubsection*{Integral for short time}
\end{singlespace}

\begin{singlespace}
Next, evaluating the integral $D_{R}(t)=\int_{0}^{t}dt'\sum_{k}|g_{k}|^{2}\bar{N_{k}}\cos\left(\omega_{k0}(t-t')\right)$
from $0$ to $t$ amounts to finding the area under the curve from
$t'=0$ to $t'=t$, as represented by the shaded area in Figures 2
and 3.

For very short time, as shown in Figure 2, the shaded area increases
(almost) linearly with $t$. This is because the curve (i.e. the integrand
as a differentiable function of $t'$) is flat in the neighborhood
of its maximum $t'=t$.
\end{singlespace}
\begin{singlespace}

\subsubsection*{Integral for long time - constant decoherence rate}
\end{singlespace}

\begin{singlespace}
For longer time, as shown in Figure 3, the shaded area stays (almost)
constant despite the increase of $t$, because the left tail of the
curve has a negligible area. Therefore, we may legitimately extend
the lower limit of the integral from $t'=0$ to $t'=-\infty$ (almost)
without changing the shaded area. (See \cite{jones} for similar discussions
on extending the limit of the integral to infinity.) In doing so,
we formally make $D_{R}(t)$ a constant:
\begin{eqnarray}
D_{R}(t) & = & \int_{0}^{t}dt'\sum_{k}|g_{k}|^{2}\bar{N_{k}}\cos\left(\omega_{k0}(t-t')\right)\nonumber \\
 & \cong & \int_{-\infty}^{t}dt'\sum_{k}|g_{k}|^{2}\bar{N_{k}}\cos\left(\omega_{k0}(t-t')\right)\nonumber \\
 & = & \int_{-\infty}^{t}dt'\sum_{k}|g_{k}|^{2}\bar{N_{k}}Re\left(e^{i\omega_{k0}\left(t-t'\right)}\right)\nonumber \\
 & = & Re\left(\int_{-\infty}^{t}dt'\sum_{k}|g_{k}|^{2}\bar{N_{k}}e^{i\omega_{k0}\left(t-t'\right)}\right)\nonumber \\
 & = & \sum_{k}|g_{k}|^{2}\bar{N_{k}}Re\left(\int_{-\infty}^{t}dt'e^{i\omega_{k0}\left(t-t'\right)}\right)\nonumber \\
 & = & \sum_{k}|g_{k}|^{2}\bar{N_{k}}Re\left(-\int_{+\infty}^{0}d\tau e^{i\omega_{k0}\tau}\right)\nonumber \\
 & = & \sum_{k}|g_{k}|^{2}\bar{N_{k}}Re\left(\int_{0}^{+\infty}d\tau e^{i\tau\omega_{k0}}\right)\nonumber \\
 & = & \sum_{k}|g_{k}|^{2}\bar{N_{k}}\pi\delta\left(-\omega_{k0}\right)\nonumber \\
 & = & \pi\sum_{k}|g_{k}|^{2}\bar{N_{k}}\delta\left(\omega_{0}-\omega_{k}\right),\label{eq:Dr new 2}
\end{eqnarray}
where in the sixth line we have made the change of variable $\tau=t-t'$
and in the eighth line we have invoked the equality $\int_{0}^{+\infty}dk\,e^{-ikx}=\pi\delta\left(x\right)-i\,Pr\frac{1}{x}$.
\cite{heitler book,jones} To facilitate further calculation of Eq.(\ref{eq:Dr new 2}),
we follow the treatment in \cite{heller book} and invoke the following
change of variable - for an arbitrary function $f(k)$: 
\begin{eqnarray}
\sum_{k}f(k) & = & \sum_{k}\Delta k\,f(k)\nonumber \\
 & = & \sum_{k}\frac{\triangle k}{\triangle\omega}\Delta\omega\,f(k)\nonumber \\
 & = & \sum_{k}\frac{\triangle k}{\triangle\omega}\Delta\omega\,f\left[k(\omega)\right]\nonumber \\
 & = & \int\rho(\omega)\,d\omega\,f\left[k(\omega)\right],
\end{eqnarray}
where $\rho(\omega)\equiv\triangle k/\triangle\omega$ is the density
of states per energy/frequency. Thus we have \cite{heller book} 
\begin{eqnarray}
D_{R}(t) & \cong & \pi\sum_{k}|g_{k}|^{2}\bar{N_{k}}\delta\left(\omega_{0}-\omega_{k}\right)\nonumber \\
 & = & \pi\int_{-\infty}^{+\infty}\rho(\omega)\,d\omega\,|g_{k(\omega)}|^{2}\bar{N}_{k(\omega)}\delta\left(\omega_{0}-\omega\right)\nonumber \\
 & = & \pi\rho(\omega_{0})\,|g_{k(\omega_{0})}|^{2}\bar{N}_{k(\omega_{0})},
\end{eqnarray}
where $k(\omega_{0})$ indicates the $k$-th bosonic mode that has
frequency $\omega=\omega_{0}$. In a similar fashion, the other prefactor
$D'_{R}(t)$ is found to be 
\begin{eqnarray}
D'_{R}(t) & \cong & \pi\rho(\omega_{0})\,|g_{k(\omega_{0})}|^{2}\left(\bar{N}_{k(\omega_{0})}+1\right).
\end{eqnarray}
Note that both prefactors become (almost) constant in this case.

Thus, in the longer time regime, the second-order equation of motion
becomes \footnote{We ignore the treatment of the unitary term here, because the main
purpose of our discussion is on the issue of decay/decoherence.}
\begin{eqnarray}
\frac{d}{dt}\rho_{t} & \cong & unitary\,term-D_{R}\left(\sigma_{-}\sigma_{+}\rho_{t}+\rho_{t}\sigma_{-}\sigma_{+}-2\sigma_{+}\rho_{t}\sigma_{-}\right)\nonumber \\
 &  & -D_{R}'\left(\sigma_{+}\sigma_{-}\rho_{t}+\rho_{t}\sigma_{+}\sigma_{-}-2\sigma_{-}\rho_{t}\sigma_{+}\right),
\end{eqnarray}
where the constant decoherence rate is $D_{R}\equiv\pi\rho(\omega_{0})\,|g_{k(\omega_{0})}|^{2}\bar{N}_{k(\omega_{0})}$
($D_{R}'\equiv\pi\rho(\omega_{0})\,|g_{k(\omega_{0})}|^{2}\left(\bar{N}_{k(\omega_{0})}+1\right)$).
This form is consistent with the Markovian master equation for a TLS
as in Eq.(3.219) of Ref.\cite{breuer book}. 

Therefore, for a TLS interacting with multiple bosonic modes at a
broad spectrum of frequencies, we have recovered the (almost) constant
decay/decoherence rate in the longer time regime. A constant decay/decoherence
rate also implies exponential decay in the relevant density matrix
element(s).

\includegraphics[scale=0.8]{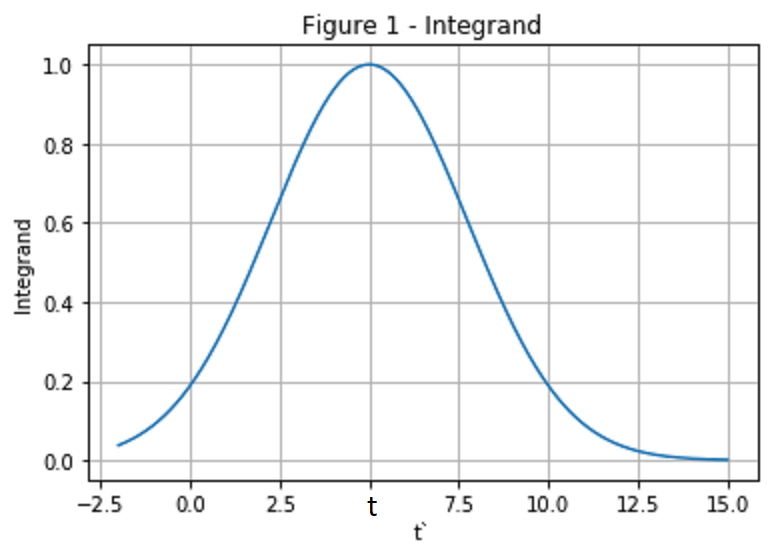}

Figure 1. The integrand $\sum_{k}|g_{k}|^{2}\bar{N_{k}}\cos\left(\omega_{k0}(t-t')\right)$
as a function of $t'$ is peaked at $t'=t$.

\includegraphics[scale=0.8]{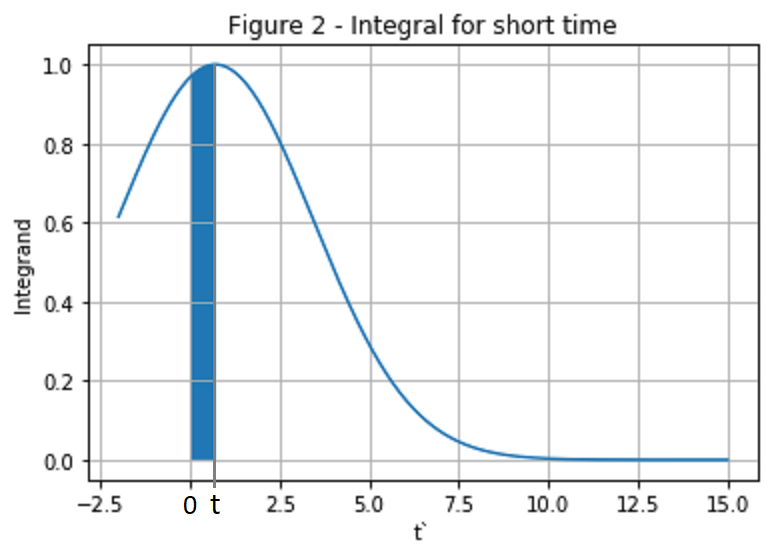}

Figure 2. For small $t$, the shaded area grows (almost) linearly
with $t$, because the curve, being a differentiable function of $t'$,
is flat in the neighborhood of its maximum $t'=t$.

\includegraphics[scale=0.8]{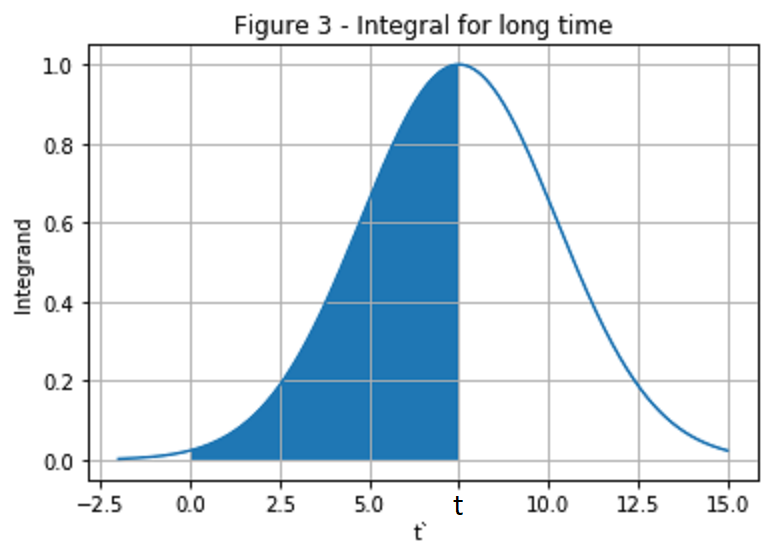}

Figure 3. For large $t$, the shaded area stays (almost) constant,
because the left tail of the integral for $t'<0$ is negligible.
\end{singlespace}

\end{document}